\documentclass[floatfix,twocolumn,jcp,aip,superscriptaddress,showpacs,nofootinbib,citeautoscript,floatfix]{revtex4-2}
\usepackage{booktabs}
\usepackage{multirow}
\usepackage{textgreek}
\usepackage{graphicx}
\usepackage{dcolumn}
\usepackage{bm}
\usepackage{flafter}
\usepackage{float}
\usepackage{lettrine}
\usepackage{amsmath}
\usepackage{amssymb}
\usepackage{color}
\usepackage{epsfig}

\graphicspath{{eps+pdf/}}
\makeatother
\DeclareUnicodeCharacter{0301}{\'{r}}
\begin{document}
\title{High pressure structural and lattice dynamics study of $\alpha$-In$_2$Se$_3$}

\author{Shiyu Feng}
\affiliation {Department of Materials Science and Engineering, Guangdong Technion-Israel Institute of Technology, Shantou 515063, China}
\affiliation{Department of Materials Science and Engineering, Technion-Israel Institute of Technology, Haifa 3200003, Israel}
\author{Baihong Sun}
\affiliation{Department of Materials Science and Engineering, Guangdong Technion-Israel Institute of Technology, Shantou 515063, China}
\affiliation{Department of Materials Science and Engineering, Technion-Israel Institute of Technology, Haifa 3200003, Israel}
\author{Wenting Lu}
\affiliation{Department of Materials Science and Engineering, Guangdong Technion-Israel Institute of Technology, Shantou 515063, China}
\affiliation{Department of Materials Science and Engineering, Technion-Israel Institute of Technology, Haifa 3200003, Israel}
\author{Haikai Zou}
\affiliation{Department of Materials Science and Engineering, Guangdong Technion-Israel Institute of Technology, Shantou 515063, China}
\affiliation{Department of Materials Science and Engineering, Technion-Israel Institute of Technology, Haifa 3200003, Israel}
\author{Chenxin Wei}
\affiliation{Department of Materials Science and Engineering, Guangdong Technion-Israel Institute of Technology, Shantou 515063, China}
\affiliation{Department of Materials Science and Engineering, Technion-Israel Institute of Technology, Haifa 3200003, Israel}
\author{Qian Zhang}
\affiliation{Department of Materials Science and Engineering, Guangdong Technion-Israel Institute of Technology, Shantou 515063, China}
\author{Bihan Wang }
\affiliation{Deutsches Elektronen-Synchrotron DESY, Notkestr. 85, Hamburg 22607, Germany}
\author{Martin Kunz}
\affiliation{Advanced Light Source,Lawrence Berkeley Laboratory,Berkeley, California 94720, USA}
\author{Hirokazu Kadobayashi}
\affiliation{SPring-8/JASRI, 1-1-1 Kouto, Sayo-gun, Sayo-cho, Hyogo 679-5198, Japan}
\author{Azkar Saeed Ahmad}
\email{azkar.ahmad@gtiit.edu.cn}
\affiliation{Department of Materials Science and Engineering, Guangdong Technion-Israel Institute of Technology, Shantou 515063, China}
\affiliation{Guangdong Provincial Key Laboratory of Materials and Technologies for Energy Conversion, Guangdong Technion-Israel Institute of Technology, Shantou 515063, China}
\author{Elad Koren}
\email{eladk@technion.ac.il}
\affiliation{Department of Materials Science and Engineering, Technion-Israel Institute of Technology, Haifa 3200003, Israel}
\author{Elissaios Stavrou}
\email{elissaios.stavrou@gtiit.edu.cn}
\affiliation{Department of Materials Science and Engineering, Guangdong Technion-Israel Institute of Technology, Shantou 515063, China}
\affiliation{Department of Materials Science and Engineering, Technion-Israel Institute of Technology, Haifa 3200003, Israel}
\affiliation{Guangdong Provincial Key Laboratory of Materials and Technologies for Energy Conversion, Guangdong Technion-Israel Institute of Technology, Shantou 515063, China}

\date{\today}%

\begin{abstract}
Layered $\alpha$-In$_2$Se$_3$  has been studied using a concomitant $in-situ$ synchrotron angle-dispersive powder x-ray diffraction and Raman spectroscopy study in a diamond anvil cell up to 60+ GPa, at room temperature. Helium, that remains fairly hydrostatic up to the highest pressure in this study, was used as the pressure-transmitting medium. The results from both experimental methods reveal a pressure-induced structural phase transition from $\alpha$-In$_2$Se$_3$ to a monoclinic $\beta'$-In$_2$Se$_3$  structure at $\approx$ 1 GPa, in agreement with previous studies. Based on our detailed measurements using both experimental techniques and $F-f$ formalism, the  $\beta'$-In$_2$Se$_3$ structure remains stable up to 45 GPa, without a clear indication of a phase transition towards the previously reported $\beta$-In$_2$Se$_3$ phase. Above this pressure,   In$_2$Se$_3$ adopts a disordered solid-solution-like orthorhombic structure, phase IV. The results are discussed in comparison with the relevant previous studies of $\alpha$-In$_2$Se$_3$ under pressure.
\end{abstract}

\maketitle

\section{Introduction}

Two-dimensional materials are ultrathin nanomaterials that are constrained in two dimensions \cite{MasBalleste2011}. They possess superior optical, mechanical, and electrical properties, such as high transparency, anisotropy, and gate tunability \cite{Mukherjee2020,Senapati2020}. Compared to conventional ferroelectric materials, their two-dimensional counterparts are more suitable for high-density storage and low-energy consumption in nanoelectronics due to their stable and outstanding ferroelectric properties \cite{Dutta2021}. $2H$ $\alpha$-In$_2$Se$_3$ is an  emerging 2D ferroelectric material and it is a III-VI semiconductor consisting of five Se-In-Se-In-Se atomic layers. It has attracted particular interest, due to the presence of robust intercoupled in-plane and out-of-plane ferroelectricity in monolayer form, that makes it highly suitable for ferroelectric-based electronics applications \cite{Ding2017,Mukherjee2022}.

At ambient conditions,  the layered rhombohedral (space group $R3m$ (160)) $\alpha$-In$_2$Se$_3$ phase is the most stable crystal structure \cite{Mukherjee2022}, and pressure is expected to significantly affect its physicochemical properties. Indeed, recent research results on other layered compounds, such as graphene \cite{Efthimiopoulos2023} and Transition Metal Dichalcogenides \cite{Nayak2014,Chi2014}, further support this hypothesis. However, to our knowledge, the overall structural behavior and the optical properties of  $\alpha$-In$_2$Se$_3$ phase under high pressure have not been systematically investigated. The majority of previous studies, using Raman spectroscopy and X-ray diffraction (XRD), were limited to 20 GPa  \cite{Liang2020,Vilaplana2018,Zhao2014} and only one XRD study \cite{Zhao2014} was conducted up to 60 GPa , albeit using a substantially non-hydrostatic pressure transmitting medium (PTM).

To address this issue, we performed a combined Raman spectroscopy and synchrotron XRD study of $\alpha$-In$_2$Se$_3$ under pressure, using helium as the pressure transmitting medium. Our results agree with the $\alpha$  $\rightarrow$ $\beta^{'}$  phase transition at $\approx$ 1 GPa, reported in previous studies \cite{Liang2020,Vilaplana2018, Zhao2014,Rasmussen2013}. However, no evidence for the transition to the previously reported  $\beta$-In$_2$Se$_3$ were found. On the other hand,  we do observe the previously reported phase transition, towards a  high-pressure phase ($IV$)\cite{Zhao2014}, above 45 GPa . We conclude that this phase is a disordered solid-solution-like orthorhombic structure, based on XRD patterns indexing and atomic volumes arguments. A detailed comparison with previous studies on $\alpha$-In$_2$Se$_3$ under pressure is also provided.

\section{Experimental methods}
The $\alpha$-In$_2$Se$_3$ samples employed in the Raman spectroscopy measurements for this study are commercially available from hq graphene BV Company in the Netherlands ($>$99.995$\%$). For X-ray diffraction (XRD) experiments, commercially available high purity ($>$99.999$\%$) $\alpha$-In$_2$Se$_3$ powder specimens  from Shanghai Aladdin Biochemical Technology Company were used.

\subsection{ High pressure studies}
Symmetric and mini-BX-80 diamond anvil cells (DAC) with diamond culets of 300-400 $\mu$m diameter were used. Between the two diamonds, a high-pressure chamber was constructed using a pre-indented Rhenium gasket with a thickness of 30-40 $\mu$m and a central hole diameter of $\approx$ 120 $\mu$m.  Powder specimens grained to $\approx$ 5 $\mu$m in average diameter were loaded, and the remaining volume was filled with Helium (He) pre-compressed to $\approx$ 2KBar (using a gas loader), acting as the pressure-transmitting medium (PTM).   He, remains fairly hydrostatic at pressures far  above its solidification pressure and up to 50+ GPa \cite{Klotz2009}.

\subsubsection{ Raman Spectroscopy}
Raman studies were performed using a custom-made confocal micro-Raman with the 532 nm line of a solid-state laser for excitation in the backscattering geometry. The laser probing spot dimension was 4 $\mu$m. Raman spectra were recorded with a spectral resolution of 2 cm$^{-1}$ using a single-stage grating spectrograph equipped with a charge-coupled device (CCD) array detector. The laser power on  specimens was kept below 1mW, to avoid any laser-induced decomposition. Ultra-low-frequency solid-state notch filters allowed us to measure Raman spectra down to 10 cm$^{-1}$ \cite{Hinton2019}. A small ruby ball was also loaded inside the sample chamber, to measure pressure using ruby luminescence \cite{Syassen2008}.

\subsubsection {X-ray Diffraction}
For the XRD measurements, small quantities of ruby and gold powder were also loaded to determine the pressure through ruby luminescence \cite{Syassen2008} and gold Equation of state (EOS) \cite{Anderson1989}, respectively.  A Dectris Pilatus3 S 1M Hybrid Photon Counting detector   was used at the Advanced Light Source, Lawrence Berkeley National Laboratory, Beamline 12.2.2. The spot size of the X-ray probing beam was focused to about 10 x 10$\mu$m  using Kirkpatrick-Baez mirrors. More details on the XRD experimental setups are given in  Kunz $et$ $al.$ \cite{Kunz2005} At SPring-8, beamline BL10XU, a Flat Panel X-ray Detector (Varex Imaging, XRD1611 CP3)  was used and the X-ray probing beam spot size was focused to about 10 x 10$\mu$m  using Kirkpatrick-Baez mirrors. More details on the SPring-8 XRD experimental setups are given in Kawaguchi‐Imada  $et$ $al$ \cite{KawaguchiImada2024}. At Beamline P02.2 at DESY, the X-ray probing beam were focused to a spot size of 2 x 2 $\mu$m  at the sample using Kirkpatrick-Baez mirrors and a PerkinElmer XRD 1621 flat-panel detector was used to collect the diffraction images of sample.

Integration of powder diffraction patterns to yield scattering intensity versus 2$\theta$ diagrams and initial analysis were performed using the DIOPTAS program \cite{Prescher2015}. Calculated XRD patterns were produced using the POWDER CELL program \cite{Kraus1996} for the corresponding crystal structures according to the EOSs determined experimentally in this study and assuming continuous Debye rings of uniform intensity. XRD patterns indexing has been performed using the DICVOL program \cite{Boultif2004} as implemented in the FullProf Suite. Le Bail refinements were performed using the $GSAS-II$ software \cite{Toby2013}.

\section{Results}
Before proceeding to the results of this study, it is useful to provide a brief summary of the previous high-pressure Raman spectroscopy and XRD studies of $\alpha$-In$_2$Se$_3$ \cite{Liang2020,Vilaplana2018,Zhao2014,Rasmussen2013}. From XRD studies \cite{Liang2020,Vilaplana2018},  it was suggested that $\alpha$-In$_2$Se$_3$ will undergo two pressure-induced phase transitions below 20 GPa: the $\alpha$ phase transforms to the $\beta^{'}$ phase at $\approx$ 0.9 GPa, which in turn converts to the $\beta$ phase at 10-12 GPa. During this process,  the coordination number of both In and Se atoms increases with pressure. Initially, for the $\alpha$ phase, the two crystallographically  nonequivalent In-1 and In-2 atoms have 4-fold and 6-fold coordination, respectively, while the Se atoms are 4-fold coordinated. After transforming to the $\beta$ phase, both In and Se atoms have the 6-fold coordination, see Fig. 7 in Ref. \onlinecite{Vilaplana2018}.

According to Zhao $et$ $al.$ \cite{Zhao2014}, In$_2$ Se$_3$ undergoes four pressure-induced transformations from ambient pressure to 59.5 GPa, based on their XRD results. At roughly 0.81 GPa, the $\alpha$-In$_2$Se$_3$ (phase $I$) transforms  to $\beta^{'}$-In$_2$Se$_3$ (phase $II$), and at around 5 GPa this phase $II$ transforms to  $\beta$-In$_2$Se$_3$ (phase $III$). Then, at about 20.6 GPa, there is an evolution from phase $III$ to $III^{'}$, similar to an isostructural phase transition. At approximately 32.1 GPa, phase $III^{'}$ finally transforms into a body-centered cubic-type structure (phase $IV$) that is stable up to, at least, 59.5 GPa.

Likewise,  two of the previous Raman spectroscopy studies indicated  a $\alpha$  $\rightarrow$ $\beta^{'}$ $\rightarrow$ $\beta$  structural sequence under pressure \cite{Liang2020,Vilaplana2018}, in agreement with XRD results. A previous study \cite{Zhao2014}  observed a prominent negative Raman shift, accompanied by a significant increase in intensity, of one of the Raman modes in the range of 170-210 $cm^{-1}$ for $\beta$-In$_2$Se$_3$ (phase III) below about 20 GPa.  However, this observation contradicts all previously reported and present results, raising questions about the origin of this effect.

We note that in the aforementioned studies, a methanol-ethanol mixture \cite{Vilaplana2018,Liang2020} or silicone oil \cite{Zhao2014} was used as the PTM. It is well established, that  both these PTMs become substantially nonhydrostatic above 12 GPa \cite{Klotz2009}, see Fig. S1. Non-hydrostatic conditions may be problematic, as it is impossible to determine the exact pressure during measurements and also they promote phase transitions to occur at lower pressures \cite{Zhang2023}. Furthermore, the same material may undergo different phase transitions under hydrostatic and nonhydrostatic conditions \cite{Gunka2021, BarredaArgueeso2013}. The above mentioned issues with respect to PTMs, was our main motivation to use He as PTM that is considered the optimal PTM, even in its solid state above 12.1 GPa at 300 K \cite{Vos1990}.

\subsection{Raman spectroscopy}
In Fig. S2, the Raman spectrum of In$_2$Se$_3$  at ambient pressure can be attributed to the $\alpha$ phase according to Ref. \onlinecite{Lewandowska2001}. According to group theory (see also table I for Wyckoff positions), the  Raman active zone center modes are $\widetilde{\Gamma}_{\alpha} $= $5A_1$+$5E$. From the 10 expected modes, the Raman peaks at 104 $cm^{-1}$, 182 $cm^{-1}$ and 203 $cm^{-1}$ are assigned to the A$_1$ symmetry, while the peak at 27 $cm^{-1}$ was assigned to the E symmetry \cite{Lewandowska2001}. Figure 1 shows selected high-pressure Raman scattering spectra in different  pressure ranges ((a)-(c)), from ambient pressure to 59.8 GPa.  Clear spectral changes can be observed above 1.44 GPa, signaling the transition to the  $\beta^{'}$ phase, in agreement with previous studies \cite{Liang2020,Vilaplana2018}. No obvious spectroscopic changes are observed up to 40 GPa,  $i.e.$ no indication for a subsequent transition from the $\beta'$ to the $\beta$ phase was observed. This is further supported by the pressure dependence of the Raman active modes in Fig. 2.

\begin{figure}[h]
    \centering
    \includegraphics[width=\linewidth]{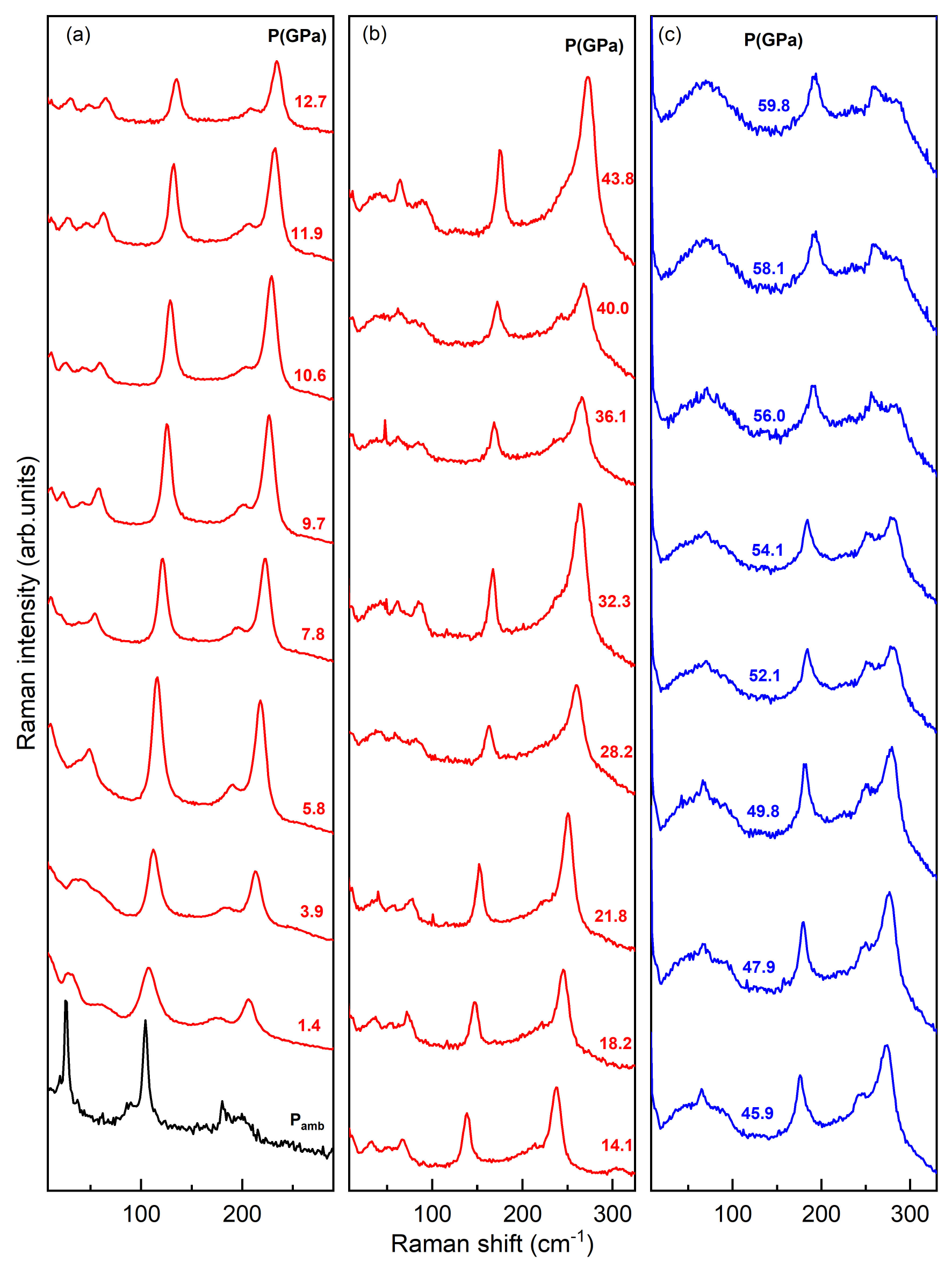}
    \caption{Selected high pressure Raman scattering spectra at different pressure ranges. a) from ambient pressure to 12.7GPa, b) from 14.1GPa to 43.8GPa and c) from 45.9GPa to 59.8GPa.}
\end{figure}

According to group theory, the  Raman active zone center modes for the $\beta' $ and $\beta$ phases are, $\widetilde{\Gamma}_{\beta^{'}}$= $4A_g+2B_g$ and $\widetilde{\Gamma}_{\beta}$ = $2A_{1g}+2E_g$, respectively.  Thus, a transition to the $\beta$ phase shall be accompanied (in principle) by a reduction of the observed Raman modes. However, in our study, we observe a continuity of both the frequencies and the number of Raman modes. It is plausible to assume that because the Raman system used in Ref. \onlinecite{Vilaplana2018} precluded  measurements bellow 50 cm$^{-1}$, an apparent decrease of the number of Raman active modes was observed. In contrast, the ability of our system to record Raman modes up to 10 cm$^{-1}$ close to the Rayleigh, gives us confidence that there is no spectral discontinuity above 12 GPa and up to 40 GPa.

\begin{figure}[h]
    \centering
    \includegraphics[width=\linewidth]{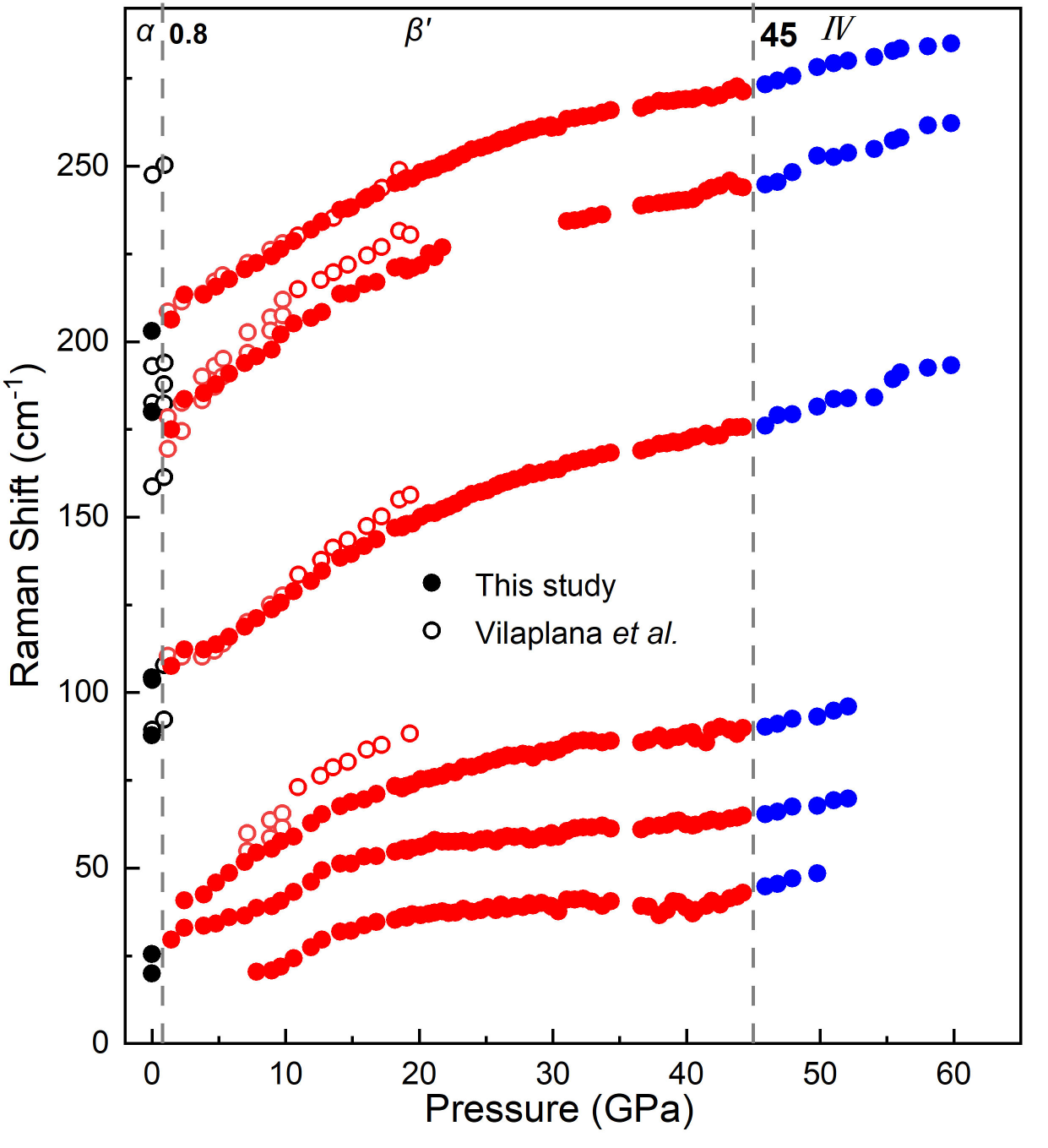}
    \caption{Pressure dependence of the Raman-active modes frequencies, according to this study (solid symbols) and from Vilaplana $et$ $al.$ \cite{Vilaplana2018} (open symbols). The $\alpha$, $\beta^{'}$,  and IV phases are indicated  with black, red,  and blue respectively. The vertical dashed lines indicate the critical pressures of the  phase transitions.}
\end{figure}

We also note, a slight discrepancy, of the Raman mode frequencies between our study and Ref. \onlinecite{Vilaplana2018}   in the 10-20 GPa pressure range. That is, our data consistently show lower frequencies than Ref.\onlinecite{Vilaplana2018}. As previously mentioned, 4:1 meth.-eth.  used as PTM in Ref. \onlinecite{Vilaplana2018}, introduce non-hydrostatic conditions above  10 GPa. Assuming a standard experimental setup, $i.e.$ pressure marker placed further form the center, while Raman signal from specimen is probed close to the center, this can be easily understood based on the developed pressure gradient under nonhydrostatic conditions, with higher ``real'' pressure than the one determined by the marker. Indeed, a slight adjustment of the frequencies of Ref. \onlinecite{Vilaplana2018} to higher pressures above 10 GPa, resulting to an overlap between the two datasets.

Above 45 GPa, although the number and the frequencies of the observed Raman modes agree with the ones of $\beta^{'}$-In$_2$Se$_3$, clear spectroscopic changes can be observed. Indeed,  a gradual inversion of the relative intensities of the high-frequency Raman modes at $\approx$  245 and 275 cm$^{-1}$ is apparent, see Fig. 1(c). We tentatively assign this change to a phase transition towards  phase IV, following Ref.  \onlinecite{Zhao2014}, with a broad pressure range of coexistence between the $\beta^{'}$-In$_2$Se$_3$ and IV phases.

During decompression,  (see Figs. S2-S4), we observed the sequential transformation of In$_2$Se$_3$  to the $\alpha$ phase with minima hysteresis.  Thus, all pressure-induced phase transitions are  reversible at room temperature, consistent with previous studies\cite{Vilaplana2018,Zhao2014}. A detailed comparison between the Raman spectrum of $\alpha$-In$_2$Se$_3$ after full pressure release with the one  before compression (Fig. S2) reveals that, except for the much lower intensity of the peak at 104$cm^{-1}$, all expected active Raman modes (see Ref. \onlinecite{Lewandowska2001}) are present in the spectrum of the released specimen.  An apparent increase of the Raman modes width  is indicative of a remaining  pressure-induced disorder.

\subsection{X-ray diffraction}
XRD patterns of In$_2$ Se$_3$ at selected pressures upon pressure increase are plotted in Figure 3. A clear pressure-induced phase transition was observed at $\approx$ 1 GPa, in agreement with the Raman results of this study and previous XRD studies \cite{Vilaplana2018, Zhao2014} and the relevant patterns above 1 GPa can be indexed with the previously reported $\beta$'-In$_2$Se$_3$ phase. From the relevant Le Bail refinements of the XRD patterns of this study (see Fig. S5), the lattice parameters and the cell volume for the  In$_2$Se$_3$  phases were determined and are plotted in Figs. 4 and 5 and listed in Table I.

\begin{figure}[h]
    \centering
    \includegraphics[width=\linewidth]{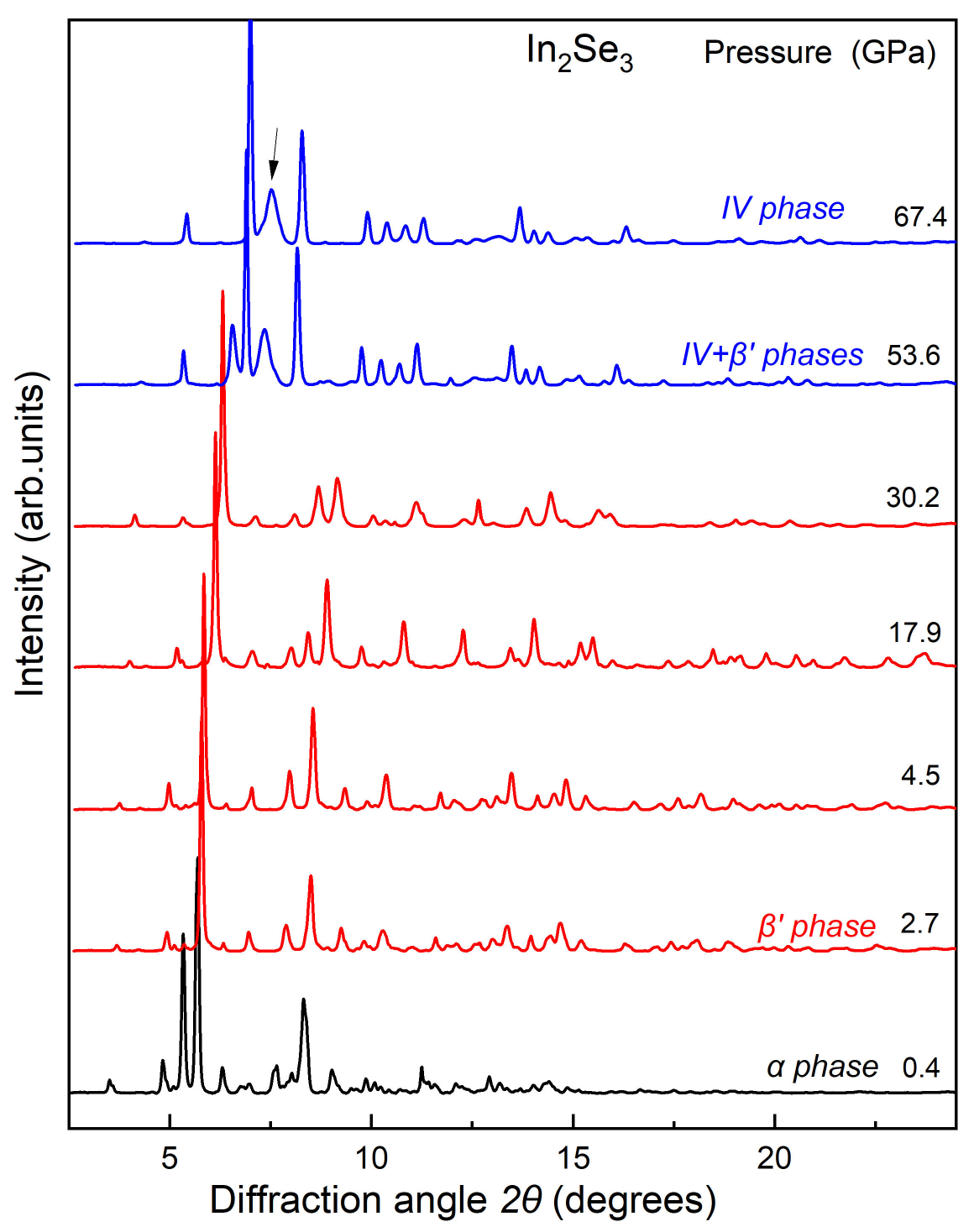}
    \caption{Selected XRD patterns of In$_2$Se$_3$ with increasing pressure. The $\alpha$-In$_2$Se$_3$, $\beta$'-In$_2$Se$_3$, and IV-In$_2$Se$_3$ phases are denoted with black, red and  blue colors, respectively. The arrow points towards a broad amorphous-like diffraction feature, see text for details. The X-ray wavelength is $\lambda$=0.2904\r{A}}
\end{figure}

\begin{table*}[tb]
\centering
\footnotesize
\caption{Structural parameters of $\alpha$-, $\beta'$- and IV-type phases of  In$_2$Se$_3$ at selected pressures as determined in this study: space group (SG), formula units in the unit cell Z, lattice parameters,  cell volume/Z, bulk modulus B and its pressure derivative B' (as determined by fitting a 3$^{rd}$ order Birch-Murnaghan  EOS \cite{Birch1947}  to the  experimental data) at the  onset pressure and  Wyckoff positions (WP).}
\medskip
\begin{ruledtabular}
\begin{tabular}{llllllllll}
P(GPa)& SG& Z & $a$({\AA}) & $b$({\AA}) & $c$({\AA}) & $V_{pfu}$({\AA}$^3$)& B(GPa)&  B' & WP \\\hline
0.4& \emph{R3m}&3&3.994&3.994&28.397&130.77&&& In:2x(3a), Se:3x(3a) \\\hline
4.5& \emph{C2$/$m}&2&6.721&3.890&9.140&115.55& 49.2(3)&5.3(12)& In:(4i), Se:(4i)+(2a)\\
33.3&\emph{C2$/$m}&2&6.153&3.600&8.338&89.764&&& In:(4i), Se:(4i)+(2a) \\\hline
62.8& &3/5 &3.776&3.778&3.084&73.3&221(5)& 8.2(15)& \\
 \end{tabular}
\end{ruledtabular}
\end{table*}

Above 12 GPa, although the XRD patterns could be indexed  with the $\beta$-In$_2$Se$_3$  phase, no clear appearance or disappearance of Bragg peaks was observed.  In fact, XRD patterns above 12 GPa can also be indexed using the  $\beta$'-In$_2$Se$_3$ phase. According to Ref. \onlinecite{Vilaplana2018}, the $\beta'$ $\rightarrow$ $\beta$ transition is characterized by a symmetrization, resulting to a decrease of the observed Bragg peaks, without volume discontinuities, pointing towards a second order transition. Indeed, no volume discontinuities were observed in this study, that could be related to the $\beta'$ $\rightarrow$ $\beta$ transition, see Fig. 5. 

We also note the difficulty on establishing a definite critical pressure for this phase transition, with a large difference  between Ref. \onlinecite{Vilaplana2018} (12 GPa) and Ref.\onlinecite{Zhao2014} (5 GPa), although both pressures are inside the hydrostaticity limit of both PTMs used in these studies. A closer look on the space groups of $\beta'$ ($C2/m$, \#12) and  $\beta$ ($R-3m$, \#166) phases, reveals that they hold a direct group-subgroup relation, and this further supports the second-order nature of this possible transition. Finally, we note that in Ref. \onlinecite{Rasmussen2013} a direct $\alpha$ $\rightarrow$ $\beta$ was reported, based on the lower number of observed Bragg peaks, and this further highlights the challenge on distinguishing  between the $\beta'$ and $\beta$ from XRD alone. Thus, it is plausible to conclude that the indexing of Bragg peaks with the $\beta$ phase is directly related with the resolution of the measured XRD patterns. Nevertheless, for completeness, and most important for direct comparison with Ref. \onlinecite{Vilaplana2018}, in Fig. S6 we plot the lattice parameters as a function of pressure by adopting the $\beta$ phase above 12 GPa.

\begin{figure}[h]
    \centering
    \includegraphics[width=\linewidth]{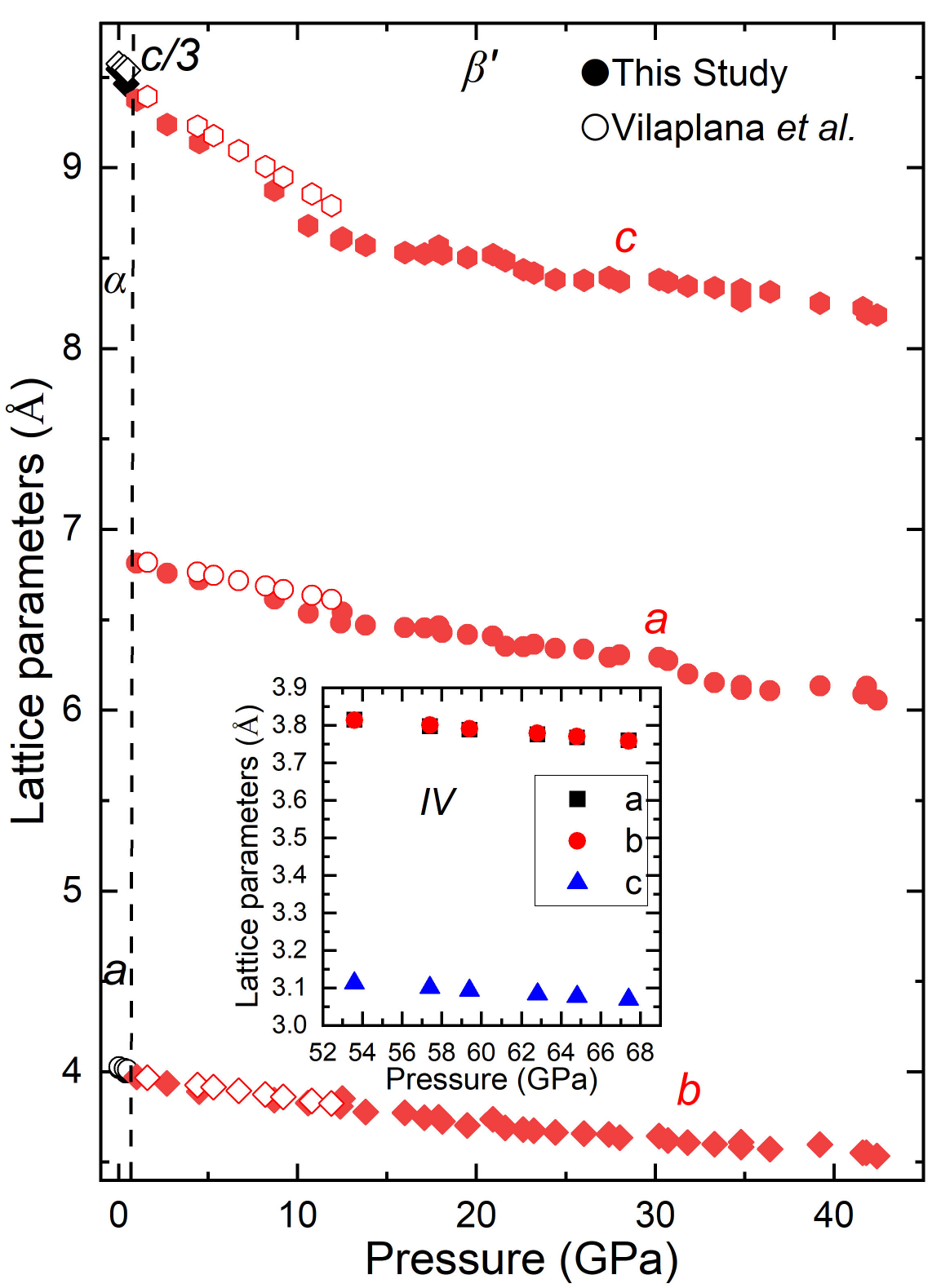}
    \caption{Pressure dependence of the lattice parameters of In$_2$Se$_3$  according to this study  (solid symbols)  and from  Vilaplana  $et$ $al.$ \cite{Vilaplana2018} (open symbols). The inset shows the lattice parameters for the  IV phase. The vertical dashed line indicates the critical pressures of the  $\alpha$ $\rightarrow$ $\beta'$ phase transition.}
\end{figure}

\begin{figure}[h]
    \centering
    \includegraphics[width=\linewidth]{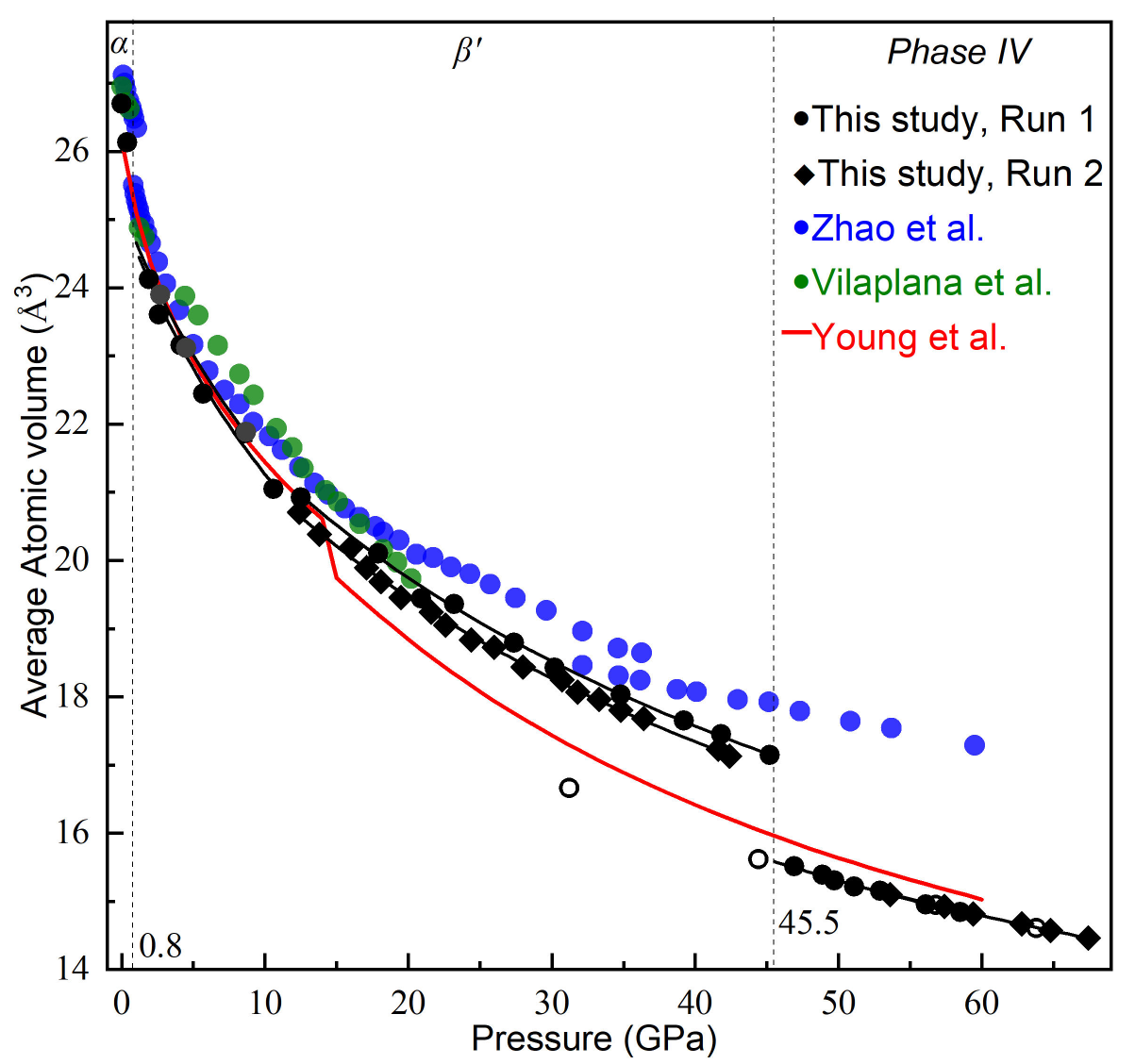}
    \caption{Pressure dependence of the average atomic volume according to this study, black solid symbols upon  pressure increase and black open symbols upon pressure release. The results from  Zhao  $et$ $al$ \cite{Zhao2014} (blue solid symbols) and Vilaplana  $et$ $al.$ \cite{Vilaplana2018} (green solid symbols) are also plotted. The DFT results according to Young $et$ $al.$ \cite{Young2016} are plotted with a red line. The vertical dashed lines indicate the critical pressures of the  phase transitions.}
\end{figure}

Above 45 GPa, the XRD patterns clearly indicate a phase transition, as evident by the appearance and disappearance of Bragg peaks, towards phase IV, in agreement with Ref. \onlinecite{Zhao2014}. The higher critical pressure in this study (45 GPa) than in Ref. \onlinecite{Zhao2014} (31 GPa), is anticipated based on the difference in the level of hydrostaticity. Moreover, the overall shape (position and intensity) of the observed XRD patterns in this study are in agreement with Ref.\onlinecite{Zhao2014}, although a much higher  number of Bragg peaks (15 vs 5) are resolved in this study.    $\beta'$-In$_2$Se$_3$ (or $\beta$-In$_2$Se$_3$, see discussion), continues to coexist with phase IV up to, at least, 54 GPa, while only phase IV can be observed at 67.4 GPa, that is the highest pressure of this XRD study. We note that both the coexistence of phases for, at least, 10 GPa and the critical pressure for the phase transition, are in agreement with the Raman spectroscopy results of this study, see Fig. 1(c). Moreover, a broad, amorphous-like feature starts to develop, see the arrow in Fig. 3. According to Zhao $et$ $al.$,   the IV structure was attributed (based on the indexing of 5 observed Bragg peaks) to a cubic defect Th$_3$P$_4$-type BCC-like structure. In this structure, Se atoms occupy the 16c WP, while In atoms occupy the 12c WP, albeit with a 0.88 occupancy, resulting to Z=5.3 and a 2/3 ratio. This structure can index all observed Bragg peaks in the previous study and all, except one low angle peak, Bragg peaks of this study  with relatively good agreement between expected and observed intensities. We postpone the evaluation of the validity of this structural assignment for the discussion section of our manuscript.

Upon pressure release, phase IV remains stable up to, at least, 32 GPa, see Fig. S7. Thus, showing considerable hysteresis  for the inverse transition to the $\beta'$-In$_2$Se$_3$ phase. This is further supported by the considerable frequency-gap (between corresponding frequencies upon increasing and decreasing pressure) observed at the same pressure range around 30 GPa for few Raman modes upon pressure release, see Fig. S4. Upon full pressure release, the XRD pattern can be indexed with the $\alpha$-In$_2$Se$_3$, albeit with considerable increase of the width of the Bragg peaks width, in agrement with the results from the Raman spectroscopy measurements.

\section{Discussion}
Both the XRD and Raman spectroscopy data of this study clearly indicate the absence of a first-order phase transition between 1 GPa and up to 45 GPa, $i.e.$ a $\beta'$ $\rightarrow$ $\beta$ transition.  In order to further explore the possibility of a subtle structural modification, we performed an additional analysis of the XRD results by plotting the PV data using the finite-strain $F-f$ formalism \cite{Birch1947}.  The $F-f$ EOS is ideal for probing subtle structural changes that relax the built-up stress of unit cells. Application of the $F-f$ EOS model reveals that pressure-dependent stress exhibits a linear response to the applied strain up to 45 GPa within the established errors; see Fig. 6. Thus, there is no indication of a pressure or strain-induced modification of the $\beta$'-In$_2$Se$_3$.  Consequently, we conclude that there is no, even a subtle, structural modification of the $\beta'$-In$_2$Se$_3$ phase up to 45 GPa.

\begin{figure}[h]
    \centering
    \includegraphics[width=\linewidth]{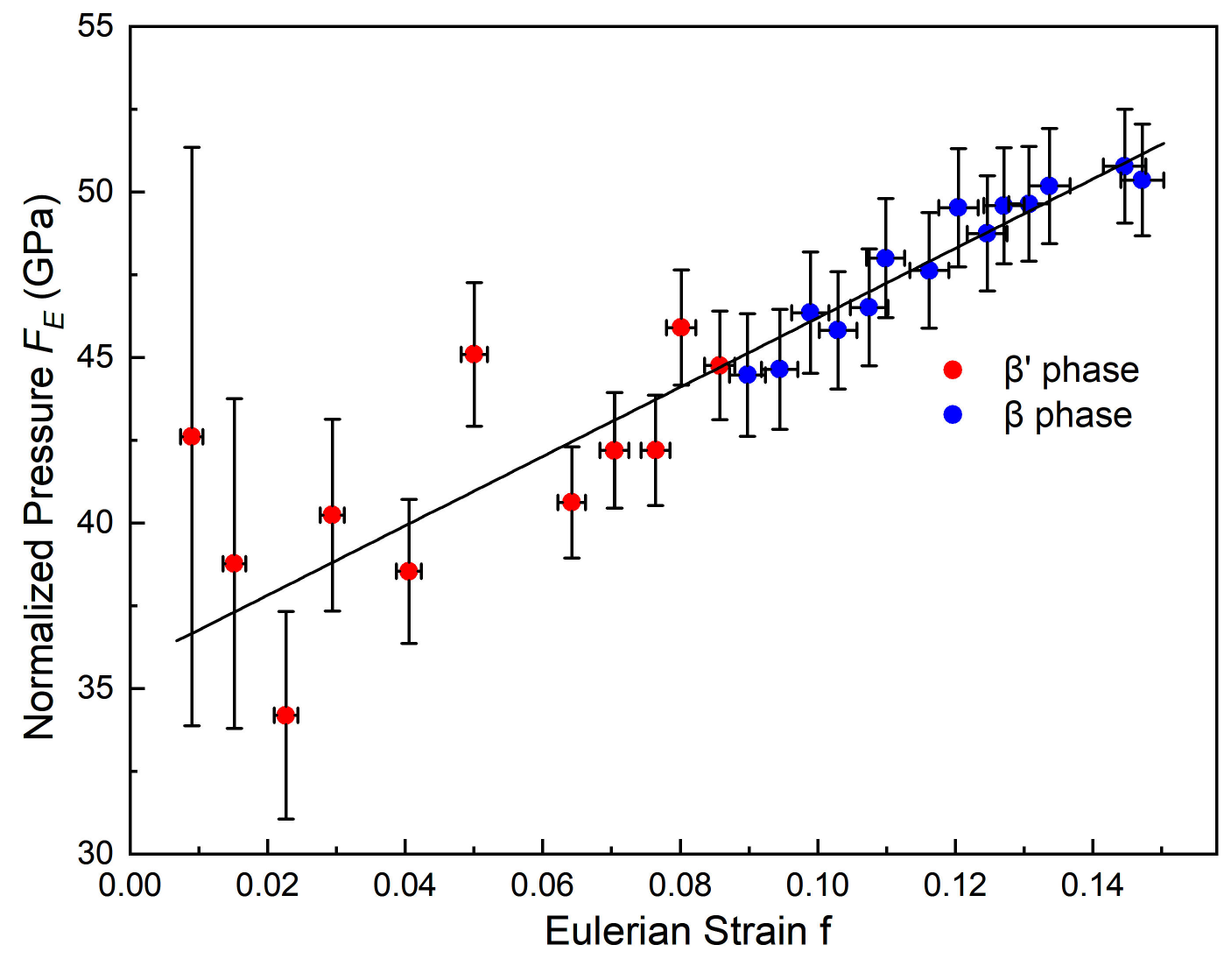}
    \caption{F- f EOS plot of the P-V data. The red and blue symbols are the P-V data of this study indexed (see text for details)  with the previously proposed $\beta'$ and $\beta$ phases, respectively. The black line is a linear fit to the F- f data.}
\end{figure}

As mentioned in the previous section, the IV structure above 45 GPa, was attributed by Zhao $et al.$   to a cubic defect Th$_3$P$_4$-type BCC-like structure. However, this structure is counterintuitive: a) it is well known that pressure promotes compact structures. The overall increase of the coordination number \cite{Stavrou2016} and the formation host-guest structures \cite{McMahon2006} under pressure  being two well-known manifestations of this general intuition and b) The resulted volume/Z for this structure is much higher than the weighted sum (2*In+3*Se) of the elemental volumes at the same pressure according to Young $et$ {} $al.$, see Fig.5 , something that points towards compositional instability.

For this reason, we performed an independent indexing, using more than 15 observed Bragg peaks. Surprisingly, the observed Bragg peaks can be well indexed (with a 3-fold higher figure of merit (FOM) than the cubic structure) with a low-volume orthorhombic structure. Most importantly, a low intensity peak at 4.3 deg cannot be indexed with the cubic defect Th$_3$P$_4$-type structure, while it was indexed with the orthorhombic structure. The volume of this orthorhombic structure (43.52 \AA$^3$ at 67 GPa) points towards a crystal structure with 3 atoms per primitive unit cell. Based on the In$_2$Se$_3$ expected stoichiometry, this structure can only be explained by the formation of a solid solution with an occupancy of 0.4 and 0.6 of the relevant crystallographic positions by In and Se, respectively. The formation of a solid solution is qualitatively in agreement with the case of Bi$_2$Ti$_3$ under pressure \cite{Zhu2011,Einaga2011}, that has practically identical structure with In$_2$Se$_3$ at ambient conditions. Indeed, Bi$_2$Te$_3$ forms a simple BCC solid solution above 20 GPa.  The higher atomic numbers of Bi and Te compared to In and Se can explain the higher pressure needed for the formation of a solid solution in the case of In$_2$Se$_3$. Unfortunately, we were not able to find a previously reported relevant structural type (oP3) that fits the unit cell used for indexing, something that points towards a new topology. Further studies, including $ab- initio$ theoretical calculations, are needed to elucidate the exact structure of phase IV.

Finally, we tentatively assign the broad (amorphous-like) XRD peak (at $\approx$ 7.5 deg, see Fig. 3) that appears in the XRD patterns of the IV phase, to a partial amorphization of In$_2$Se$_3$ above this pressure. This might be  related to the formation of the disorder solid-solution IV phase. We have also considered the possibility of a, even partial, decomposition of In$_2$Se$_3$ at this pressure. However, the continuity of the observed Raman peaks throughout the $\beta'$ $\rightarrow$ $IV$ transition; see Figs. 2 and 3, rules out this possibility.

\section{Conclusion}
According to the results of both Raman spectroscopy and the XRD techniques used in this study, $\alpha$-In$_2$Se$_3$  undergoes a phase transition at $\approx$ 1 GPa towards the previously reported $\beta'$-In$_2$Se$_3$ phase. In contrast to the previous studies, we conclude that this phase remains stable up to 45 GPa, at a pressure that another pressure-induced phase transition is observed towards phase IV. Based on XRD patterns indexing and atomic volumes arguments, we  conclude that phase IV is  a disordered solid-solution-like orthorhombic structure.

\begin{acknowledgments}
S.F. and C.W.  acknowledges support from  the Graduate scholarships of the Guangdong Provincial Key Laboratory of Materials and Technologies for Energy Conversion.
The work performed at GTIIT was supported by funding from the Guangdong Technion Israel Institute of Technology and the Guangdong Provincial Key Laboratory of Materials and Technologies for Energy Conversion, MATEC (No. MATEC2022KF001).
Beamline 12.2.2 at the Advanced Light Source is a DOE Office of Science User Facility under contract no. DE-AC02-05CH11231.
Part of the synchrotron radiation experiments were performed at BL10XU  of SPring-8 with the approval of the Japan Synchrotron Radiation Research Institute (JASRI) (Proposal Nos. 2024A1096 and 2024B1182).
We also acknowledge DESY (Hamburg, Germany), a member of the Helmholtz Association HGF, for the provision of experimental facilities. Parts of this research were carried out at PETRA III beamline P02.2.
\end{acknowledgments}

\section*{AUTHOR DECLARATIONS} 

\subsection*{Conflict of Interest} 
The authors have no conflicts to disclose.

\section*{DATA AVAILABILITY}
The data that support the findings of this study are available from the corresponding author upon reasonable request.

\bibliography{In2Se3}

\begin{thebibliography}{34}%
\makeatletter
\providecommand \@ifxundefined [1]{%
 \@ifx{#1\undefined}
}%
\providecommand \@ifnum [1]{%
 \ifnum #1\expandafter \@firstoftwo
 \else \expandafter \@secondoftwo
 \fi
}%
\providecommand \@ifx [1]{%
 \ifx #1\expandafter \@firstoftwo
 \else \expandafter \@secondoftwo
 \fi
}%
\providecommand \natexlab [1]{#1}%
\providecommand \enquote  [1]{``#1''}%
\providecommand \bibnamefont  [1]{#1}%
\providecommand \bibfnamefont [1]{#1}%
\providecommand \citenamefont [1]{#1}%
\providecommand \href@noop [0]{\@secondoftwo}%
\providecommand \href [0]{\begingroup \@sanitize@url \@href}%
\providecommand \@href[1]{\@@startlink{#1}\@@href}%
\providecommand \@@href[1]{\endgroup#1\@@endlink}%
\providecommand \@sanitize@url [0]{\catcode `\\12\catcode `\$12\catcode
  `\&12\catcode `\#12\catcode `\^12\catcode `\_12\catcode `\%12\relax}%
\providecommand \@@startlink[1]{}%
\providecommand \@@endlink[0]{}%
\providecommand \url  [0]{\begingroup\@sanitize@url \@url }%
\providecommand \@url [1]{\endgroup\@href {#1}{\urlprefix }}%
\providecommand \urlprefix  [0]{URL }%
\providecommand \Eprint [0]{\href }%
\providecommand \doibase [0]{https://doi.org/}%
\providecommand \selectlanguage [0]{\@gobble}%
\providecommand \bibinfo  [0]{\@secondoftwo}%
\providecommand \bibfield  [0]{\@secondoftwo}%
\providecommand \translation [1]{[#1]}%
\providecommand \BibitemOpen [0]{}%
\providecommand \bibitemStop [0]{}%
\providecommand \bibitemNoStop [0]{.\EOS\space}%
\providecommand \EOS [0]{\spacefactor3000\relax}%
\providecommand \BibitemShut  [1]{\csname bibitem#1\endcsname}%
\let\auto@bib@innerbib\@empty
\bibitem [{\citenamefont {Mas-Balleste}\ \emph {et~al.}(2011)\citenamefont
  {Mas-Balleste}, \citenamefont {Gomez-Navarro}, \citenamefont
  {Gomez-Herrero},\ and\ \citenamefont {Zamora}}]{MasBalleste2011}%
  \BibitemOpen
  \bibfield  {author} {\bibinfo {author} {\bibfnamefont {R.}~\bibnamefont
  {Mas-Balleste}}, \bibinfo {author} {\bibfnamefont {C.}~\bibnamefont
  {Gomez-Navarro}}, \bibinfo {author} {\bibfnamefont {J.}~\bibnamefont
  {Gomez-Herrero}},\ and\ \bibinfo {author} {\bibfnamefont {F.}~\bibnamefont
  {Zamora}},\ }\bibfield  {title} {\enquote {\bibinfo {title} {2{D} materials:
  to graphene and beyond},}\ }\href@noop {} {\bibfield  {journal} {\bibinfo
  {journal} {Nanoscale}\ }\textbf {\bibinfo {volume} {3}},\ \bibinfo {pages}
  {20--30} (\bibinfo {year} {2011})}\BibitemShut {NoStop}%
\bibitem [{\citenamefont {Mukherjee}\ \emph {et~al.}(2020)\citenamefont
  {Mukherjee}, \citenamefont {Dutta}, \citenamefont {Mohapatra}, \citenamefont
  {Dezanashvili}, \citenamefont {Ismach},\ and\ \citenamefont
  {Koren}}]{Mukherjee2020}%
  \BibitemOpen
  \bibfield  {author} {\bibinfo {author} {\bibfnamefont {S.}~\bibnamefont
  {Mukherjee}}, \bibinfo {author} {\bibfnamefont {D.}~\bibnamefont {Dutta}},
  \bibinfo {author} {\bibfnamefont {P.~K.}\ \bibnamefont {Mohapatra}}, \bibinfo
  {author} {\bibfnamefont {L.}~\bibnamefont {Dezanashvili}}, \bibinfo {author}
  {\bibfnamefont {A.}~\bibnamefont {Ismach}},\ and\ \bibinfo {author}
  {\bibfnamefont {E.}~\bibnamefont {Koren}},\ }\bibfield  {title} {\enquote
  {\bibinfo {title} {Scalable integration of coplanar heterojunction monolithic
  devices on two-dimensional {In}$_2${Se}$_3$},}\ }\href@noop {} {\bibfield
  {journal} {\bibinfo  {journal} {ACS nano}\ }\textbf {\bibinfo {volume}
  {14}},\ \bibinfo {pages} {17543--17553} (\bibinfo {year} {2020})}\BibitemShut
  {NoStop}%
\bibitem [{\citenamefont {Senapati}\ and\ \citenamefont
  {Maiti}(2020)}]{Senapati2020}%
  \BibitemOpen
  \bibfield  {author} {\bibinfo {author} {\bibfnamefont {S.}~\bibnamefont
  {Senapati}}\ and\ \bibinfo {author} {\bibfnamefont {P.}~\bibnamefont
  {Maiti}},\ }\enquote {\bibinfo {title} {Emerging bio-applications of
  two-dimensional nanoheterostructure materials},}\ in\ \href@noop {} {\emph
  {\bibinfo {booktitle} {2D nanoscale heterostructured materials}}}\ (\bibinfo
  {publisher} {Elsevier},\ \bibinfo {year} {2020})\ pp.\ \bibinfo {pages}
  {243--255}\BibitemShut {NoStop}%
\bibitem [{\citenamefont {Dutta}\ \emph {et~al.}(2021)\citenamefont {Dutta},
  \citenamefont {Mukherjee}, \citenamefont {Uzhansky},\ and\ \citenamefont
  {Koren}}]{Dutta2021}%
  \BibitemOpen
  \bibfield  {author} {\bibinfo {author} {\bibfnamefont {D.}~\bibnamefont
  {Dutta}}, \bibinfo {author} {\bibfnamefont {S.}~\bibnamefont {Mukherjee}},
  \bibinfo {author} {\bibfnamefont {M.}~\bibnamefont {Uzhansky}},\ and\
  \bibinfo {author} {\bibfnamefont {E.}~\bibnamefont {Koren}},\ }\bibfield
  {title} {\enquote {\bibinfo {title} {Cross-field optoelectronic modulation
  via inter-coupled ferroelectricity in 2{D} {In}$_2${Se}$_3$},}\ }\href@noop
  {} {\bibfield  {journal} {\bibinfo  {journal} {npj 2D Materials and
  Applications}\ }\textbf {\bibinfo {volume} {5}},\ \bibinfo {pages} {81}
  (\bibinfo {year} {2021})}\BibitemShut {NoStop}%
\bibitem [{\citenamefont {Ding}\ \emph {et~al.}(2017)\citenamefont {Ding},
  \citenamefont {Zhu}, \citenamefont {Wang}, \citenamefont {Gao}, \citenamefont
  {Xiao}, \citenamefont {Gu}, \citenamefont {Zhang},\ and\ \citenamefont
  {Zhu}}]{Ding2017}%
  \BibitemOpen
  \bibfield  {author} {\bibinfo {author} {\bibfnamefont {W.}~\bibnamefont
  {Ding}}, \bibinfo {author} {\bibfnamefont {J.}~\bibnamefont {Zhu}}, \bibinfo
  {author} {\bibfnamefont {Z.}~\bibnamefont {Wang}}, \bibinfo {author}
  {\bibfnamefont {Y.}~\bibnamefont {Gao}}, \bibinfo {author} {\bibfnamefont
  {D.}~\bibnamefont {Xiao}}, \bibinfo {author} {\bibfnamefont {Y.}~\bibnamefont
  {Gu}}, \bibinfo {author} {\bibfnamefont {Z.}~\bibnamefont {Zhang}},\ and\
  \bibinfo {author} {\bibfnamefont {W.}~\bibnamefont {Zhu}},\ }\bibfield
  {title} {\enquote {\bibinfo {title} {Prediction of intrinsic two-dimensional
  ferroelectrics in {In}$_2${Se}$_3$ and other {III}2-{VI}3 van der waals
  materials},}\ }\href@noop {} {\bibfield  {journal} {\bibinfo  {journal} {Nat.
  commun.}\ }\textbf {\bibinfo {volume} {8}},\ \bibinfo {pages} {14956}
  (\bibinfo {year} {2017})}\BibitemShut {NoStop}%
\bibitem [{\citenamefont {Mukherjee}\ and\ \citenamefont
  {Koren}(2022)}]{Mukherjee2022}%
  \BibitemOpen
  \bibfield  {author} {\bibinfo {author} {\bibfnamefont {S.}~\bibnamefont
  {Mukherjee}}\ and\ \bibinfo {author} {\bibfnamefont {E.}~\bibnamefont
  {Koren}},\ }\bibfield  {title} {\enquote {\bibinfo {title} {Indium selenide
  ({In}$_2${Se}$_3$) – an emerging van‐der‐waals material for
  photodetection and non‐volatile memory applications},}\ }\href@noop {}
  {\bibfield  {journal} {\bibinfo  {journal} {Isr. J. Chem.}\ }\textbf
  {\bibinfo {volume} {62}} (\bibinfo {year} {2022})}\BibitemShut {NoStop}%
\bibitem [{\citenamefont {Efthimiopoulos}\ \emph {et~al.}(2023)\citenamefont
  {Efthimiopoulos}, \citenamefont {Stavrou}, \citenamefont {Umemoto},
  \citenamefont {Mayanna}, \citenamefont {Torode}, \citenamefont {Smith},
  \citenamefont {Chariton}, \citenamefont {Prakapenka}, \citenamefont
  {Goncharov},\ and\ \citenamefont {Wang}}]{Efthimiopoulos2023}%
  \BibitemOpen
  \bibfield  {author} {\bibinfo {author} {\bibfnamefont {I.}~\bibnamefont
  {Efthimiopoulos}}, \bibinfo {author} {\bibfnamefont {E.}~\bibnamefont
  {Stavrou}}, \bibinfo {author} {\bibfnamefont {K.}~\bibnamefont {Umemoto}},
  \bibinfo {author} {\bibfnamefont {S.}~\bibnamefont {Mayanna}}, \bibinfo
  {author} {\bibfnamefont {A.}~\bibnamefont {Torode}}, \bibinfo {author}
  {\bibfnamefont {J.~S.}\ \bibnamefont {Smith}}, \bibinfo {author}
  {\bibfnamefont {S.}~\bibnamefont {Chariton}}, \bibinfo {author}
  {\bibfnamefont {V.~B.}\ \bibnamefont {Prakapenka}}, \bibinfo {author}
  {\bibfnamefont {A.~F.}\ \bibnamefont {Goncharov}},\ and\ \bibinfo {author}
  {\bibfnamefont {Y.}~\bibnamefont {Wang}},\ }\bibfield  {title} {\enquote
  {\bibinfo {title} {High-pressure phase of cold-compressed bulk graphite and
  graphene nanoplatelets},}\ }\href
  {https://doi.org/10.1103/PhysRevB.107.184102} {\bibfield  {journal} {\bibinfo
   {journal} {Phys. Rev. B}\ }\textbf {\bibinfo {volume} {107}},\ \bibinfo
  {pages} {184102} (\bibinfo {year} {2023})}\BibitemShut {NoStop}%
\bibitem [{\citenamefont {Nayak}\ \emph {et~al.}(2014)\citenamefont {Nayak},
  \citenamefont {Bhattacharyya}, \citenamefont {Zhu}, \citenamefont {Liu},
  \citenamefont {Wu}, \citenamefont {Pandey}, \citenamefont {Jin},
  \citenamefont {Singh}, \citenamefont {Akinwande},\ and\ \citenamefont
  {Lin}}]{Nayak2014}%
  \BibitemOpen
  \bibfield  {author} {\bibinfo {author} {\bibfnamefont {A.~P.}\ \bibnamefont
  {Nayak}}, \bibinfo {author} {\bibfnamefont {S.}~\bibnamefont
  {Bhattacharyya}}, \bibinfo {author} {\bibfnamefont {J.}~\bibnamefont {Zhu}},
  \bibinfo {author} {\bibfnamefont {J.}~\bibnamefont {Liu}}, \bibinfo {author}
  {\bibfnamefont {X.}~\bibnamefont {Wu}}, \bibinfo {author} {\bibfnamefont
  {T.}~\bibnamefont {Pandey}}, \bibinfo {author} {\bibfnamefont
  {C.}~\bibnamefont {Jin}}, \bibinfo {author} {\bibfnamefont {A.~K.}\
  \bibnamefont {Singh}}, \bibinfo {author} {\bibfnamefont {D.}~\bibnamefont
  {Akinwande}},\ and\ \bibinfo {author} {\bibfnamefont {J.-F.}\ \bibnamefont
  {Lin}},\ }\bibfield  {title} {\enquote {\bibinfo {title} {Pressure-induced
  semiconducting to metallic transition in multilayered molybdenum
  disulphide},}\ }\href {https://doi.org/10.1038/ncomms4731} {\bibfield
  {journal} {\bibinfo  {journal} {Nature Communications}\ }\textbf {\bibinfo
  {volume} {5}},\ \bibinfo {pages} {3731} (\bibinfo {year} {2014})}\BibitemShut
  {NoStop}%
\bibitem [{\citenamefont {Chi}\ \emph {et~al.}(2014)\citenamefont {Chi},
  \citenamefont {Zhao}, \citenamefont {Zhang}, \citenamefont {Goncharov},
  \citenamefont {Lobanov}, \citenamefont {Kagayama}, \citenamefont {Sakata},\
  and\ \citenamefont {Chen}}]{Chi2014}%
  \BibitemOpen
  \bibfield  {author} {\bibinfo {author} {\bibfnamefont {Z.-H.}\ \bibnamefont
  {Chi}}, \bibinfo {author} {\bibfnamefont {X.-M.}\ \bibnamefont {Zhao}},
  \bibinfo {author} {\bibfnamefont {H.}~\bibnamefont {Zhang}}, \bibinfo
  {author} {\bibfnamefont {A.~F.}\ \bibnamefont {Goncharov}}, \bibinfo {author}
  {\bibfnamefont {S.~S.}\ \bibnamefont {Lobanov}}, \bibinfo {author}
  {\bibfnamefont {T.}~\bibnamefont {Kagayama}}, \bibinfo {author}
  {\bibfnamefont {M.}~\bibnamefont {Sakata}},\ and\ \bibinfo {author}
  {\bibfnamefont {X.-J.}\ \bibnamefont {Chen}},\ }\bibfield  {title} {\enquote
  {\bibinfo {title} {Pressure-induced metallization of molybdenum disulfide},}\
  }\href {https://doi.org/10.1103/PhysRevLett.113.036802} {\bibfield  {journal}
  {\bibinfo  {journal} {Phys. Rev. Lett.}\ }\textbf {\bibinfo {volume} {113}},\
  \bibinfo {pages} {036802} (\bibinfo {year} {2014})}\BibitemShut {NoStop}%
\bibitem [{\citenamefont {Liang}\ \emph {et~al.}(2020)\citenamefont {Liang},
  \citenamefont {Jin}, \citenamefont {Zhang},\ and\ \citenamefont
  {Chen}}]{Liang2020}%
  \BibitemOpen
  \bibfield  {author} {\bibinfo {author} {\bibfnamefont {J.}~\bibnamefont
  {Liang}}, \bibinfo {author} {\bibfnamefont {H.}~\bibnamefont {Jin}}, \bibinfo
  {author} {\bibfnamefont {J.}~\bibnamefont {Zhang}},\ and\ \bibinfo {author}
  {\bibfnamefont {X.}~\bibnamefont {Chen}},\ }\bibfield  {title} {\enquote
  {\bibinfo {title} {Structural evolution of β’-{In}$_2${Se}$_3$under
  pressure},}\ }\href {https://doi.org/10.1088/1742-6596/1622/1/012027}
  {\bibfield  {journal} {\bibinfo  {journal} {J. Phys.: Conf. Ser.}\ }\textbf
  {\bibinfo {volume} {1622}},\ \bibinfo {pages} {012027} (\bibinfo {year}
  {2020})}\BibitemShut {NoStop}%
\bibitem [{\citenamefont {Vilaplana}\ \emph {et~al.}(2018)\citenamefont
  {Vilaplana}, \citenamefont {Parra}, \citenamefont {Jorge-Montero},
  \citenamefont {Rodríguez-Hernández}, \citenamefont {Munoz}, \citenamefont
  {Errandonea}, \citenamefont {Segura},\ and\ \citenamefont
  {Manjón}}]{Vilaplana2018}%
  \BibitemOpen
  \bibfield  {author} {\bibinfo {author} {\bibfnamefont {R.}~\bibnamefont
  {Vilaplana}}, \bibinfo {author} {\bibfnamefont {S.~G.}\ \bibnamefont
  {Parra}}, \bibinfo {author} {\bibfnamefont {A.}~\bibnamefont
  {Jorge-Montero}}, \bibinfo {author} {\bibfnamefont {P.}~\bibnamefont
  {Rodríguez-Hernández}}, \bibinfo {author} {\bibfnamefont {A.}~\bibnamefont
  {Munoz}}, \bibinfo {author} {\bibfnamefont {D.}~\bibnamefont {Errandonea}},
  \bibinfo {author} {\bibfnamefont {A.}~\bibnamefont {Segura}},\ and\ \bibinfo
  {author} {\bibfnamefont {F.~J.}\ \bibnamefont {Manjón}},\ }\bibfield
  {title} {\enquote {\bibinfo {title} {Experimental and theoretical studies on
  α-{In}$_2${Se}$_3$ at high pressure},}\ }\href@noop {} {\bibfield  {journal}
  {\bibinfo  {journal} {Inorg. Chem.}\ }\textbf {\bibinfo {volume} {57}},\
  \bibinfo {pages} {8241--8252} (\bibinfo {year} {2018})}\BibitemShut {NoStop}%
\bibitem [{\citenamefont {Zhao}\ and\ \citenamefont {Yang}(2014)}]{Zhao2014}%
  \BibitemOpen
  \bibfield  {author} {\bibinfo {author} {\bibfnamefont {J.}~\bibnamefont
  {Zhao}}\ and\ \bibinfo {author} {\bibfnamefont {L.}~\bibnamefont {Yang}},\
  }\bibfield  {title} {\enquote {\bibinfo {title} {Structure evolutions and
  metallic transitions in {In}$_2${Se}$_3$ under high pressure},}\ }\href@noop
  {} {\bibfield  {journal} {\bibinfo  {journal} {J. Phys. Chem. C}\ }\textbf
  {\bibinfo {volume} {118}},\ \bibinfo {pages} {5445--5452} (\bibinfo {year}
  {2014})}\BibitemShut {NoStop}%
\bibitem [{\citenamefont {Rasmussen}\ \emph {et~al.}(2013)\citenamefont
  {Rasmussen}, \citenamefont {Teklemichael}, \citenamefont {Mafi},
  \citenamefont {Gu},\ and\ \citenamefont {McCluskey}}]{Rasmussen2013}%
  \BibitemOpen
  \bibfield  {author} {\bibinfo {author} {\bibfnamefont {A.~M.}\ \bibnamefont
  {Rasmussen}}, \bibinfo {author} {\bibfnamefont {S.~T.}\ \bibnamefont
  {Teklemichael}}, \bibinfo {author} {\bibfnamefont {E.}~\bibnamefont {Mafi}},
  \bibinfo {author} {\bibfnamefont {Y.}~\bibnamefont {Gu}},\ and\ \bibinfo
  {author} {\bibfnamefont {M.~D.}\ \bibnamefont {McCluskey}},\ }\bibfield
  {title} {\enquote {\bibinfo {title} {Pressure-induced phase transformation of
  {In}$_2${Se}$_3$},}\ }\href {https://doi.org/10.1063/1.4792313} {\bibfield
  {journal} {\bibinfo  {journal} {Appl. Phys. Lett.}\ }\textbf {\bibinfo
  {volume} {102}},\ \bibinfo {pages} {062105} (\bibinfo {year}
  {2013})}\BibitemShut {NoStop}%
\bibitem [{\citenamefont {Klotz}\ \emph {et~al.}(2009)\citenamefont {Klotz},
  \citenamefont {Chervin}, \citenamefont {Munsch},\ and\ \citenamefont
  {Le~Marchand}}]{Klotz2009}%
  \BibitemOpen
  \bibfield  {author} {\bibinfo {author} {\bibfnamefont {S.}~\bibnamefont
  {Klotz}}, \bibinfo {author} {\bibfnamefont {J.}~\bibnamefont {Chervin}},
  \bibinfo {author} {\bibfnamefont {P.}~\bibnamefont {Munsch}},\ and\ \bibinfo
  {author} {\bibfnamefont {G.}~\bibnamefont {Le~Marchand}},\ }\bibfield
  {title} {\enquote {\bibinfo {title} {Hydrostatic limits of 11 pressure
  transmitting media},}\ }\href@noop {} {\bibfield  {journal} {\bibinfo
  {journal} {J. Phys. D: Appl. Phys.}\ }\textbf {\bibinfo {volume} {42}},\
  \bibinfo {pages} {075413} (\bibinfo {year} {2009})}\BibitemShut {NoStop}%
\bibitem [{\citenamefont {Hinton}\ \emph {et~al.}(2019)\citenamefont {Hinton},
  \citenamefont {Clarke}, \citenamefont {Steele}, \citenamefont {Kuo},
  \citenamefont {Greenberg}, \citenamefont {Prakapenka}, \citenamefont {Kunz},
  \citenamefont {Kroonblawd},\ and\ \citenamefont {Stavrou}}]{Hinton2019}%
  \BibitemOpen
  \bibfield  {author} {\bibinfo {author} {\bibfnamefont {J.~K.}\ \bibnamefont
  {Hinton}}, \bibinfo {author} {\bibfnamefont {S.~M.}\ \bibnamefont {Clarke}},
  \bibinfo {author} {\bibfnamefont {B.~A.}\ \bibnamefont {Steele}}, \bibinfo
  {author} {\bibfnamefont {I.-F.~W.}\ \bibnamefont {Kuo}}, \bibinfo {author}
  {\bibfnamefont {E.}~\bibnamefont {Greenberg}}, \bibinfo {author}
  {\bibfnamefont {V.~B.}\ \bibnamefont {Prakapenka}}, \bibinfo {author}
  {\bibfnamefont {M.}~\bibnamefont {Kunz}}, \bibinfo {author} {\bibfnamefont
  {M.~P.}\ \bibnamefont {Kroonblawd}},\ and\ \bibinfo {author} {\bibfnamefont
  {E.}~\bibnamefont {Stavrou}},\ }\bibfield  {title} {\enquote {\bibinfo
  {title} {Effects of pressure on the structure and lattice dynamics of
  α-glycine: a combined experimental and theoretical study},}\ }\href
  {https://doi.org/10.1039/C8CE02123F} {\bibfield  {journal} {\bibinfo
  {journal} {CrystEngComm}\ }\textbf {\bibinfo {volume} {21}},\ \bibinfo
  {pages} {4457--4464} (\bibinfo {year} {2019})}\BibitemShut {NoStop}%
\bibitem [{\citenamefont {Syassen}(2008)}]{Syassen2008}%
  \BibitemOpen
  \bibfield  {author} {\bibinfo {author} {\bibfnamefont {K.}~\bibnamefont
  {Syassen}},\ }\bibfield  {title} {\enquote {\bibinfo {title} {Ruby under
  pressure},}\ }\href@noop {} {\bibfield  {journal} {\bibinfo  {journal} {High
  Pressure Research}\ }\textbf {\bibinfo {volume} {28}},\ \bibinfo {pages}
  {75--126} (\bibinfo {year} {2008})}\BibitemShut {NoStop}%
\bibitem [{\citenamefont {Anderson}, \citenamefont {Isaak},\ and\ \citenamefont
  {Yamamoto}(1989)}]{Anderson1989}%
  \BibitemOpen
  \bibfield  {author} {\bibinfo {author} {\bibfnamefont {O.~L.}\ \bibnamefont
  {Anderson}}, \bibinfo {author} {\bibfnamefont {D.~G.}\ \bibnamefont
  {Isaak}},\ and\ \bibinfo {author} {\bibfnamefont {S.}~\bibnamefont
  {Yamamoto}},\ }\bibfield  {title} {\enquote {\bibinfo {title} {{Anharmonicity
  and the equation of state for gold}},}\ }\href
  {https://doi.org/10.1063/1.342969} {\bibfield  {journal} {\bibinfo  {journal}
  {J. Appl. Phys.}\ }\textbf {\bibinfo {volume} {65}},\ \bibinfo {pages}
  {1534--1543} (\bibinfo {year} {1989})}\BibitemShut {NoStop}%
\bibitem [{\citenamefont {Kunz}\ \emph {et~al.}(2005)\citenamefont {Kunz},
  \citenamefont {MacDowell}, \citenamefont {Caldwell}, \citenamefont {Cambie},
  \citenamefont {Celestre}, \citenamefont {Domning}, \citenamefont {Duarte},
  \citenamefont {Gleason}, \citenamefont {Glossinger}, \citenamefont {Kelez},
  \citenamefont {Plate}, \citenamefont {Yu}, \citenamefont {Zaug},
  \citenamefont {Padmore}, \citenamefont {Jeanloz}, \citenamefont
  {Alivisatos},\ and\ \citenamefont {Clark}}]{Kunz2005}%
  \BibitemOpen
  \bibfield  {author} {\bibinfo {author} {\bibfnamefont {M.}~\bibnamefont
  {Kunz}}, \bibinfo {author} {\bibfnamefont {A.}~\bibnamefont {MacDowell}},
  \bibinfo {author} {\bibfnamefont {W.}~\bibnamefont {Caldwell}}, \bibinfo
  {author} {\bibfnamefont {D.}~\bibnamefont {Cambie}}, \bibinfo {author}
  {\bibfnamefont {R.}~\bibnamefont {Celestre}}, \bibinfo {author}
  {\bibfnamefont {E.}~\bibnamefont {Domning}}, \bibinfo {author} {\bibfnamefont
  {R.}~\bibnamefont {Duarte}}, \bibinfo {author} {\bibfnamefont
  {A.}~\bibnamefont {Gleason}}, \bibinfo {author} {\bibfnamefont
  {J.}~\bibnamefont {Glossinger}}, \bibinfo {author} {\bibfnamefont
  {N.}~\bibnamefont {Kelez}}, \bibinfo {author} {\bibfnamefont
  {D.}~\bibnamefont {Plate}}, \bibinfo {author} {\bibfnamefont
  {T.}~\bibnamefont {Yu}}, \bibinfo {author} {\bibfnamefont {J.}~\bibnamefont
  {Zaug}}, \bibinfo {author} {\bibfnamefont {H.}~\bibnamefont {Padmore}},
  \bibinfo {author} {\bibfnamefont {R.}~\bibnamefont {Jeanloz}}, \bibinfo
  {author} {\bibfnamefont {A.}~\bibnamefont {Alivisatos}},\ and\ \bibinfo
  {author} {\bibfnamefont {S.}~\bibnamefont {Clark}},\ }\bibfield  {title}
  {\enquote {\bibinfo {title} {A beamline for high-pressure studies at the
  advanced light source with a superconducting bending magnet as the source},}\
  }\href@noop {} {\bibfield  {journal} {\bibinfo  {journal} {J. Synchrotron
  Radiat.}\ }\textbf {\bibinfo {volume} {12}},\ \bibinfo {pages} {650}
  (\bibinfo {year} {2005})}\BibitemShut {NoStop}%
\bibitem [{\citenamefont {Kawaguchi-Imada}\ \emph {et~al.}(2024)\citenamefont
  {Kawaguchi-Imada}, \citenamefont {Sinmyo}, \citenamefont {Ohta},
  \citenamefont {Kawaguchi},\ and\ \citenamefont
  {Kobayashi}}]{KawaguchiImada2024}%
  \BibitemOpen
  \bibfield  {author} {\bibinfo {author} {\bibfnamefont {S.}~\bibnamefont
  {Kawaguchi-Imada}}, \bibinfo {author} {\bibfnamefont {R.}~\bibnamefont
  {Sinmyo}}, \bibinfo {author} {\bibfnamefont {K.}~\bibnamefont {Ohta}},
  \bibinfo {author} {\bibfnamefont {S.}~\bibnamefont {Kawaguchi}},\ and\
  \bibinfo {author} {\bibfnamefont {T.}~\bibnamefont {Kobayashi}},\ }\bibfield
  {title} {\enquote {\bibinfo {title} {Submillisecond in situ x-ray diffraction
  measurement system with changing temperature and pressure using diamond anvil
  cells at {BL10XU/SPring-8}.}}\ }\href@noop {} {\bibfield  {journal} {\bibinfo
   {journal} {J. Synchrotron Radiat.}\ }\textbf {\bibinfo {volume} {31}},\
  \bibinfo {pages} {343--354} (\bibinfo {year} {2024})}\BibitemShut {NoStop}%
\bibitem [{\citenamefont {Prescher}\ and\ \citenamefont
  {Prakapenka}(2015)}]{Prescher2015}%
  \BibitemOpen
  \bibfield  {author} {\bibinfo {author} {\bibfnamefont {C.}~\bibnamefont
  {Prescher}}\ and\ \bibinfo {author} {\bibfnamefont {V.~B.}\ \bibnamefont
  {Prakapenka}},\ }\bibfield  {title} {\enquote {\bibinfo {title} {Dioptas: a
  program for reduction of two-dimensional x-ray diffraction data and data
  exploration},}\ }\href@noop {} {\bibfield  {journal} {\bibinfo  {journal}
  {High Pres. Res.}\ }\textbf {\bibinfo {volume} {35}},\ \bibinfo {pages}
  {223--230} (\bibinfo {year} {2015})}\BibitemShut {NoStop}%
\bibitem [{\citenamefont {Kraus}\ and\ \citenamefont
  {Nolze}(1996)}]{Kraus1996}%
  \BibitemOpen
  \bibfield  {author} {\bibinfo {author} {\bibfnamefont {W.}~\bibnamefont
  {Kraus}}\ and\ \bibinfo {author} {\bibfnamefont {G.}~\bibnamefont {Nolze}},\
  }\bibfield  {title} {\enquote {\bibinfo {title} {{{\it POWDER CELL} {--} a
  program for the representation and manipulation of crystal structures and
  calculation of the resulting X-ray powder patterns}},}\ }\href@noop {}
  {\bibfield  {journal} {\bibinfo  {journal} {J. Appl. Crystallogr.}\ }\textbf
  {\bibinfo {volume} {29}},\ \bibinfo {pages} {301--303} (\bibinfo {year}
  {1996})}\BibitemShut {NoStop}%
\bibitem [{\citenamefont {Boultif}\ and\ \citenamefont
  {Lou\"{e}r}(2004)}]{Boultif2004}%
  \BibitemOpen
  \bibfield  {author} {\bibinfo {author} {\bibfnamefont {A.}~\bibnamefont
  {Boultif}}\ and\ \bibinfo {author} {\bibfnamefont {D.}~\bibnamefont
  {Lou\"{e}r}},\ }\bibfield  {title} {\enquote {\bibinfo {title} {Powder
  pattern indexing with the dichotomy method},}\ }\href@noop {} {\bibfield
  {journal} {\bibinfo  {journal} {J. Appl. Crystallogr.}\ }\textbf {\bibinfo
  {volume} {37}},\ \bibinfo {pages} {724--731} (\bibinfo {year}
  {2004})}\BibitemShut {NoStop}%
\bibitem [{\citenamefont {Toby}\ and\ \citenamefont
  {Von~Dreele}(2013)}]{Toby2013}%
  \BibitemOpen
  \bibfield  {author} {\bibinfo {author} {\bibfnamefont {B.~H.}\ \bibnamefont
  {Toby}}\ and\ \bibinfo {author} {\bibfnamefont {R.~B.}\ \bibnamefont
  {Von~Dreele}},\ }\bibfield  {title} {\enquote {\bibinfo {title} {{{\it
  GSAS-II}: the genesis of a modern open-source all purpose crystallography
  software package}},}\ }\href {https://doi.org/10.1107/S0021889813003531}
  {\bibfield  {journal} {\bibinfo  {journal} {Journal of Applied
  Crystallography}\ }\textbf {\bibinfo {volume} {46}},\ \bibinfo {pages}
  {544--549} (\bibinfo {year} {2013})}\BibitemShut {NoStop}%
\bibitem [{\citenamefont {Zhang}\ \emph {et~al.}(2023)\citenamefont {Zhang},
  \citenamefont {Dai}, \citenamefont {Hu}, \citenamefont {Hong},\ and\
  \citenamefont {Li}}]{Zhang2023}%
  \BibitemOpen
  \bibfield  {author} {\bibinfo {author} {\bibfnamefont {X.}~\bibnamefont
  {Zhang}}, \bibinfo {author} {\bibfnamefont {L.}~\bibnamefont {Dai}}, \bibinfo
  {author} {\bibfnamefont {H.}~\bibnamefont {Hu}}, \bibinfo {author}
  {\bibfnamefont {M.}~\bibnamefont {Hong}},\ and\ \bibinfo {author}
  {\bibfnamefont {C.}~\bibnamefont {Li}},\ }\bibfield  {title} {\enquote
  {\bibinfo {title} {Pressure-induced reversible structural phase transitions
  and metallization in {GeTe} under hydrostatic and non-hydrostatic
  environments up to 22.9 {GPa}},}\ }\href@noop {} {\bibfield  {journal}
  {\bibinfo  {journal} {J. Non-Cryst. Solids}\ }\textbf {\bibinfo {volume}
  {618}},\ \bibinfo {pages} {122516} (\bibinfo {year} {2023})}\BibitemShut
  {NoStop}%
\bibitem [{\citenamefont {Guńka}\ \emph {et~al.}(2021)\citenamefont {Guńka},
  \citenamefont {Olejniczak}, \citenamefont {Fanetti}, \citenamefont {Bini},
  \citenamefont {Collings}, \citenamefont {Svitlyk},\ and\ \citenamefont
  {Dziubek}}]{Gunka2021}%
  \BibitemOpen
  \bibfield  {author} {\bibinfo {author} {\bibfnamefont {P.~A.}\ \bibnamefont
  {Guńka}}, \bibinfo {author} {\bibfnamefont {A.}~\bibnamefont {Olejniczak}},
  \bibinfo {author} {\bibfnamefont {S.}~\bibnamefont {Fanetti}}, \bibinfo
  {author} {\bibfnamefont {R.}~\bibnamefont {Bini}}, \bibinfo {author}
  {\bibfnamefont {I.~E.}\ \bibnamefont {Collings}}, \bibinfo {author}
  {\bibfnamefont {V.}~\bibnamefont {Svitlyk}},\ and\ \bibinfo {author}
  {\bibfnamefont {K.~F.}\ \bibnamefont {Dziubek}},\ }\bibfield  {title}
  {\enquote {\bibinfo {title} {Crystal structure and non‐hydrostatic
  stress‐induced phase transition of urotropine under high pressure},}\
  }\href@noop {} {\bibfield  {journal} {\bibinfo  {journal} {Chemistry–A
  European Journal}\ }\textbf {\bibinfo {volume} {27}},\ \bibinfo {pages}
  {1094--1102} (\bibinfo {year} {2021})}\BibitemShut {NoStop}%
\bibitem [{\citenamefont {Barreda-Argüeso}\ \emph {et~al.}(2013)\citenamefont
  {Barreda-Argüeso}, \citenamefont {López-Moreno}, \citenamefont
  {Sanz-Ortiz}, \citenamefont {Aguado}, \citenamefont {Valiente}, \citenamefont
  {González}, \citenamefont {Rodríguez}, \citenamefont {Romero},
  \citenamefont {Muñoz},\ and\ \citenamefont {Nataf}}]{BarredaArgueeso2013}%
  \BibitemOpen
  \bibfield  {author} {\bibinfo {author} {\bibfnamefont {J.~A.}\ \bibnamefont
  {Barreda-Argüeso}}, \bibinfo {author} {\bibfnamefont {S.}~\bibnamefont
  {López-Moreno}}, \bibinfo {author} {\bibfnamefont {M.~N.}\ \bibnamefont
  {Sanz-Ortiz}}, \bibinfo {author} {\bibfnamefont {F.}~\bibnamefont {Aguado}},
  \bibinfo {author} {\bibfnamefont {R.}~\bibnamefont {Valiente}}, \bibinfo
  {author} {\bibfnamefont {J.}~\bibnamefont {González}}, \bibinfo {author}
  {\bibfnamefont {F.}~\bibnamefont {Rodríguez}}, \bibinfo {author}
  {\bibfnamefont {A.~H.}\ \bibnamefont {Romero}}, \bibinfo {author}
  {\bibfnamefont {A.}~\bibnamefont {Muñoz}},\ and\ \bibinfo {author}
  {\bibfnamefont {L.}~\bibnamefont {Nataf}},\ }\bibfield  {title} {\enquote
  {\bibinfo {title} {Pressure-induced phase-transition sequence in {CoF}$_2$:
  An experimental and first-principles study on the crystal, vibrational, and
  electronic properties},}\ }\href@noop {} {\bibfield  {journal} {\bibinfo
  {journal} {Physical Review B}\ }\textbf {\bibinfo {volume} {88}},\ \bibinfo
  {pages} {214108} (\bibinfo {year} {2013})}\BibitemShut {NoStop}%
\bibitem [{\citenamefont {Vos}, \citenamefont {Van~Hinsberg},\ and\
  \citenamefont {Schouten}(1990)}]{Vos1990}%
  \BibitemOpen
  \bibfield  {author} {\bibinfo {author} {\bibfnamefont {W.~L.}\ \bibnamefont
  {Vos}}, \bibinfo {author} {\bibfnamefont {M.~G.}\ \bibnamefont
  {Van~Hinsberg}},\ and\ \bibinfo {author} {\bibfnamefont {J.~A.}\ \bibnamefont
  {Schouten}},\ }\bibfield  {title} {\enquote {\bibinfo {title} {High-pressure
  triple point in helium: The melting line of helium up to 240 kbar},}\
  }\href@noop {} {\bibfield  {journal} {\bibinfo  {journal} {Physical Review
  B}\ }\textbf {\bibinfo {volume} {42}},\ \bibinfo {pages} {6106} (\bibinfo
  {year} {1990})}\BibitemShut {NoStop}%
\bibitem [{\citenamefont {Lewandowska}\ \emph {et~al.}(2001)\citenamefont
  {Lewandowska}, \citenamefont {Bacewicz}, \citenamefont {Filipowicz},\ and\
  \citenamefont {Paszkowicz}}]{Lewandowska2001}%
  \BibitemOpen
  \bibfield  {author} {\bibinfo {author} {\bibfnamefont {R.}~\bibnamefont
  {Lewandowska}}, \bibinfo {author} {\bibfnamefont {R.}~\bibnamefont
  {Bacewicz}}, \bibinfo {author} {\bibfnamefont {J.}~\bibnamefont
  {Filipowicz}},\ and\ \bibinfo {author} {\bibfnamefont {W.}~\bibnamefont
  {Paszkowicz}},\ }\bibfield  {title} {\enquote {\bibinfo {title} {Raman
  scattering in α-{In}$_2${Se}$_3$ crystals},}\ }\href
  {https://doi.org/https://doi.org/10.1016/S0025-5408(01)00746-2} {\bibfield
  {journal} {\bibinfo  {journal} {Materials Research Bulletin}\ }\textbf
  {\bibinfo {volume} {36}},\ \bibinfo {pages} {2577--2583} (\bibinfo {year}
  {2001})}\BibitemShut {NoStop}%
\bibitem [{\citenamefont {Birch}(1947)}]{Birch1947}%
  \BibitemOpen
  \bibfield  {author} {\bibinfo {author} {\bibfnamefont {F.}~\bibnamefont
  {Birch}},\ }\bibfield  {title} {\enquote {\bibinfo {title} {Finite elastic
  strain of cubic crystals},}\ }\href {https://doi.org/10.1103/PhysRev.71.809}
  {\bibfield  {journal} {\bibinfo  {journal} {Phys. Rev.}\ }\textbf {\bibinfo
  {volume} {71}},\ \bibinfo {pages} {809--824} (\bibinfo {year}
  {1947})}\BibitemShut {NoStop}%
\bibitem [{\citenamefont {Young}\ \emph {et~al.}(2016)\citenamefont {Young},
  \citenamefont {Cynn}, \citenamefont {Söderlind},\ and\ \citenamefont
  {Landa}}]{Young2016}%
  \BibitemOpen
  \bibfield  {author} {\bibinfo {author} {\bibfnamefont {D.~A.}\ \bibnamefont
  {Young}}, \bibinfo {author} {\bibfnamefont {H.}~\bibnamefont {Cynn}},
  \bibinfo {author} {\bibfnamefont {P.}~\bibnamefont {Söderlind}},\ and\
  \bibinfo {author} {\bibfnamefont {A.}~\bibnamefont {Landa}},\ }\bibfield
  {title} {\enquote {\bibinfo {title} {Zero-{K}elvin compression isotherms of
  the elements 1 $\leq$ {Z} $\leq $92 to 100 {GPa}},}\ }\href
  {https://doi.org/10.1063/1.4963086} {\bibfield  {journal} {\bibinfo
  {journal} {Journal of Physical and Chemical Reference Data}\ }\textbf
  {\bibinfo {volume} {45}},\ \bibinfo {pages} {043101} (\bibinfo {year}
  {2016})}\BibitemShut {NoStop}%
\bibitem [{\citenamefont {Stavrou}\ \emph {et~al.}(2016)\citenamefont
  {Stavrou}, \citenamefont {Yao}, \citenamefont {Goncharov}, \citenamefont
  {Kon\^opkov\'a},\ and\ \citenamefont {Raptis}}]{Stavrou2016}%
  \BibitemOpen
  \bibfield  {author} {\bibinfo {author} {\bibfnamefont {E.}~\bibnamefont
  {Stavrou}}, \bibinfo {author} {\bibfnamefont {Y.}~\bibnamefont {Yao}},
  \bibinfo {author} {\bibfnamefont {A.~F.}\ \bibnamefont {Goncharov}}, \bibinfo
  {author} {\bibfnamefont {Z.}~\bibnamefont {Kon\^opkov\'a}},\ and\ \bibinfo
  {author} {\bibfnamefont {C.}~\bibnamefont {Raptis}},\ }\bibfield  {title}
  {\enquote {\bibinfo {title} {High-pressure structural study of
  {MnF}$_{2}$},}\ }\href {https://doi.org/10.1103/PhysRevB.93.054101}
  {\bibfield  {journal} {\bibinfo  {journal} {Phys. Rev. B}\ }\textbf {\bibinfo
  {volume} {93}},\ \bibinfo {pages} {054101} (\bibinfo {year}
  {2016})}\BibitemShut {NoStop}%
\bibitem [{\citenamefont {McMahon}\ \emph {et~al.}(2006)\citenamefont
  {McMahon}, \citenamefont {Nelmes}, \citenamefont {Schwarz},\ and\
  \citenamefont {Syassen}}]{McMahon2006}%
  \BibitemOpen
  \bibfield  {author} {\bibinfo {author} {\bibfnamefont {M.~I.}\ \bibnamefont
  {McMahon}}, \bibinfo {author} {\bibfnamefont {R.~J.}\ \bibnamefont {Nelmes}},
  \bibinfo {author} {\bibfnamefont {U.}~\bibnamefont {Schwarz}},\ and\ \bibinfo
  {author} {\bibfnamefont {K.}~\bibnamefont {Syassen}},\ }\bibfield  {title}
  {\enquote {\bibinfo {title} {Composite incommensurate {K-III} and a
  commensurate form: Study of a high-pressure phase of potassium},}\ }\href
  {https://doi.org/10.1103/PhysRevB.74.140102} {\bibfield  {journal} {\bibinfo
  {journal} {Phys. Rev. B}\ }\textbf {\bibinfo {volume} {74}},\ \bibinfo
  {pages} {140102} (\bibinfo {year} {2006})}\BibitemShut {NoStop}%
\bibitem [{\citenamefont {Zhu}\ \emph {et~al.}(2011)\citenamefont {Zhu},
  \citenamefont {Wang}, \citenamefont {Wang}, \citenamefont {Lv}, \citenamefont
  {Ma}, \citenamefont {Cui}, \citenamefont {Ma},\ and\ \citenamefont
  {Zou}}]{Zhu2011}%
  \BibitemOpen
  \bibfield  {author} {\bibinfo {author} {\bibfnamefont {L.}~\bibnamefont
  {Zhu}}, \bibinfo {author} {\bibfnamefont {H.}~\bibnamefont {Wang}}, \bibinfo
  {author} {\bibfnamefont {Y.}~\bibnamefont {Wang}}, \bibinfo {author}
  {\bibfnamefont {J.}~\bibnamefont {Lv}}, \bibinfo {author} {\bibfnamefont
  {Y.}~\bibnamefont {Ma}}, \bibinfo {author} {\bibfnamefont {Q.}~\bibnamefont
  {Cui}}, \bibinfo {author} {\bibfnamefont {Y.}~\bibnamefont {Ma}},\ and\
  \bibinfo {author} {\bibfnamefont {G.}~\bibnamefont {Zou}},\ }\bibfield
  {title} {\enquote {\bibinfo {title} {Substitutional {Alloy} of {Bi} and {Te}
  at {High} {Pressure}},}\ }\href
  {https://doi.org/10.1103/PhysRevLett.106.145501} {\bibfield  {journal}
  {\bibinfo  {journal} {Phys. Rev. Lett.}\ }\textbf {\bibinfo {volume} {106}},\
  \bibinfo {pages} {145501} (\bibinfo {year} {2011})}\BibitemShut {NoStop}%
\bibitem [{\citenamefont {Einaga}\ \emph {et~al.}(2011)\citenamefont {Einaga},
  \citenamefont {Ohmura}, \citenamefont {Nakayama}, \citenamefont {Ishikawa},
  \citenamefont {Yamada},\ and\ \citenamefont {Nakano}}]{Einaga2011}%
  \BibitemOpen
  \bibfield  {author} {\bibinfo {author} {\bibfnamefont {M.}~\bibnamefont
  {Einaga}}, \bibinfo {author} {\bibfnamefont {A.}~\bibnamefont {Ohmura}},
  \bibinfo {author} {\bibfnamefont {A.}~\bibnamefont {Nakayama}}, \bibinfo
  {author} {\bibfnamefont {F.}~\bibnamefont {Ishikawa}}, \bibinfo {author}
  {\bibfnamefont {Y.}~\bibnamefont {Yamada}},\ and\ \bibinfo {author}
  {\bibfnamefont {S.}~\bibnamefont {Nakano}},\ }\bibfield  {title} {\enquote
  {\bibinfo {title} {Pressure-induced phase transition of {Bi}$_2${Te}$_3$ to a
  bcc structure},}\ }\href {https://doi.org/10.1103/PhysRevB.83.092102}
  {\bibfield  {journal} {\bibinfo  {journal} {Phys. Rev. B}\ }\textbf {\bibinfo
  {volume} {83}},\ \bibinfo {pages} {092102} (\bibinfo {year}
  {2011})}\BibitemShut {NoStop}%
\end{thebibliography}%

\clearpage
\widetext
\begin{center}

\large\textbf{Supplementary material  for \textquotedblleft High pressure structural and lattice dynamics study of $\alpha$-In$_2$Se$_3$\textquotedblright}

\end{center}

\renewcommand{\thefigure}{S\arabic{figure}}
\setcounter{figure}{0}
\renewcommand{\thetable}{S\Roman{table}}

\begin{figure}[h]
    \centering
    \includegraphics[width=0.5\linewidth]{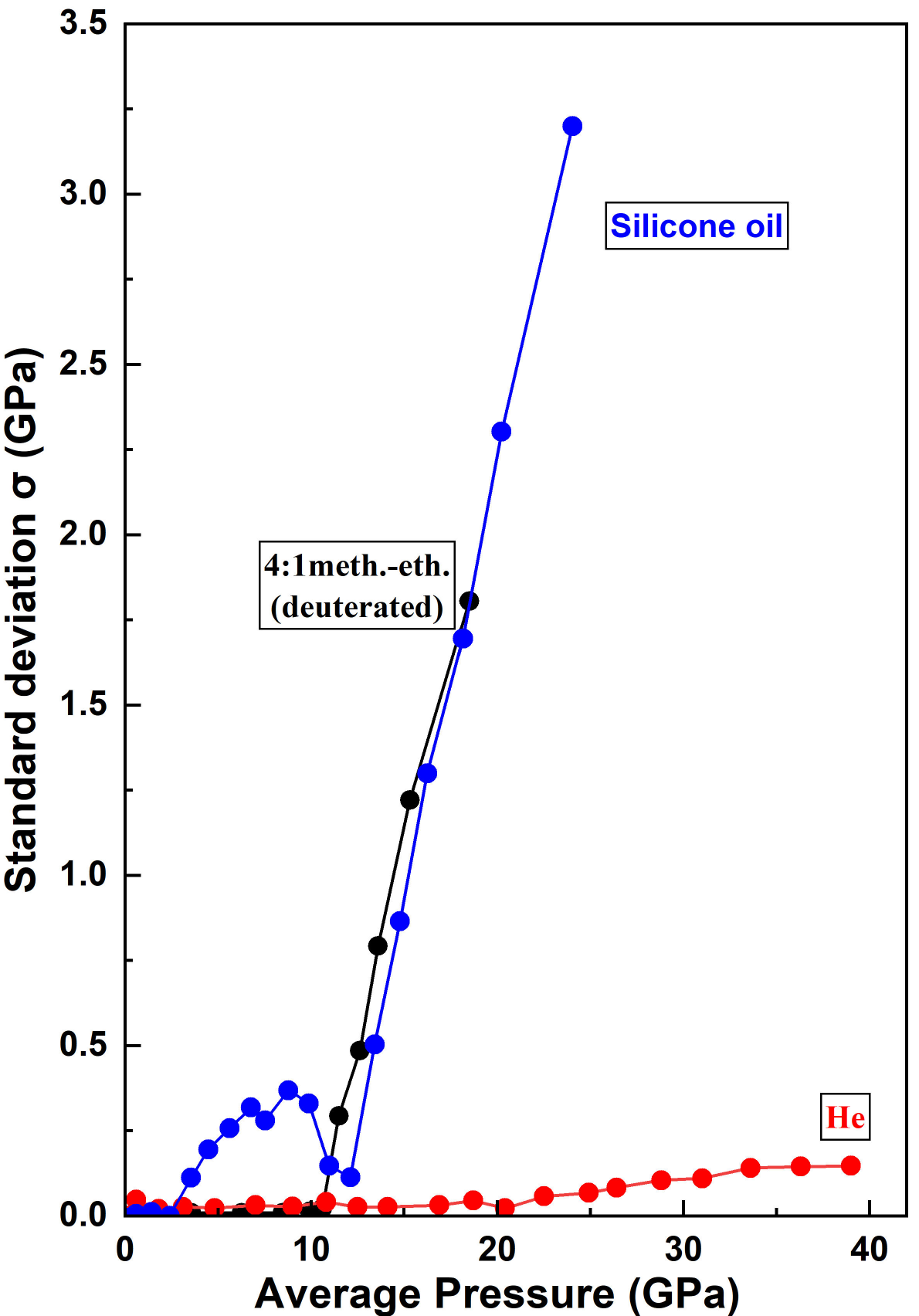}
    \caption{Comparison of pressure dependencies of the standard deviation $\sigma$ of 4:1 meth eth. (deuterated), Silicone oil and He PTMs, data are taken from cited in Ref. \onlinecite{Klotz2009}. }
\end{figure}

\begin{figure}
    \centering
    \includegraphics[width=0.7\linewidth]{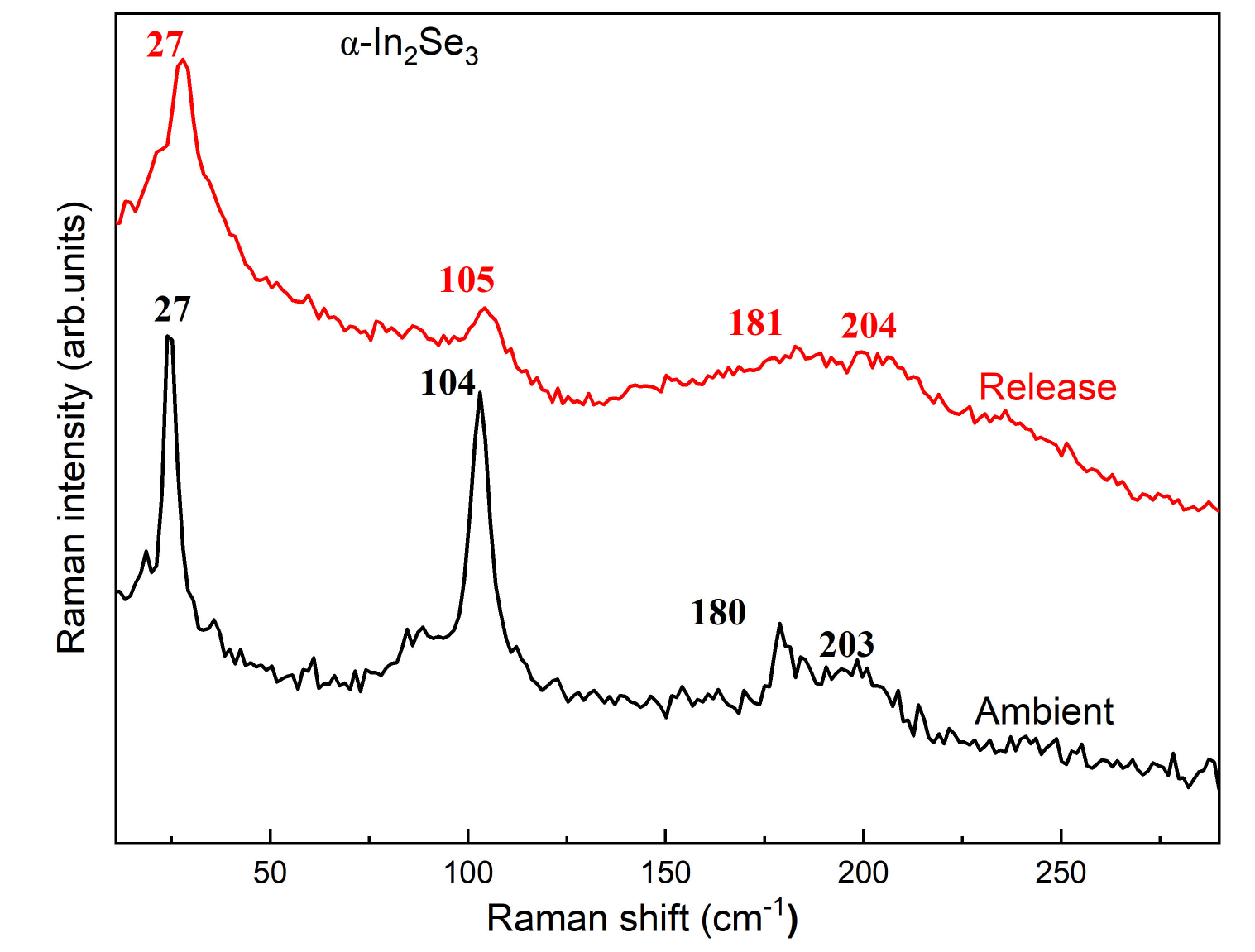}
    \caption{Raman spectra of $\alpha$-In$_2$ Se$_3$ measured at ambient pressure  before compression (black) and  after full pressure release  (red).}
\end{figure}

\begin{figure}
    \centering
    \includegraphics[width=0.5\linewidth]{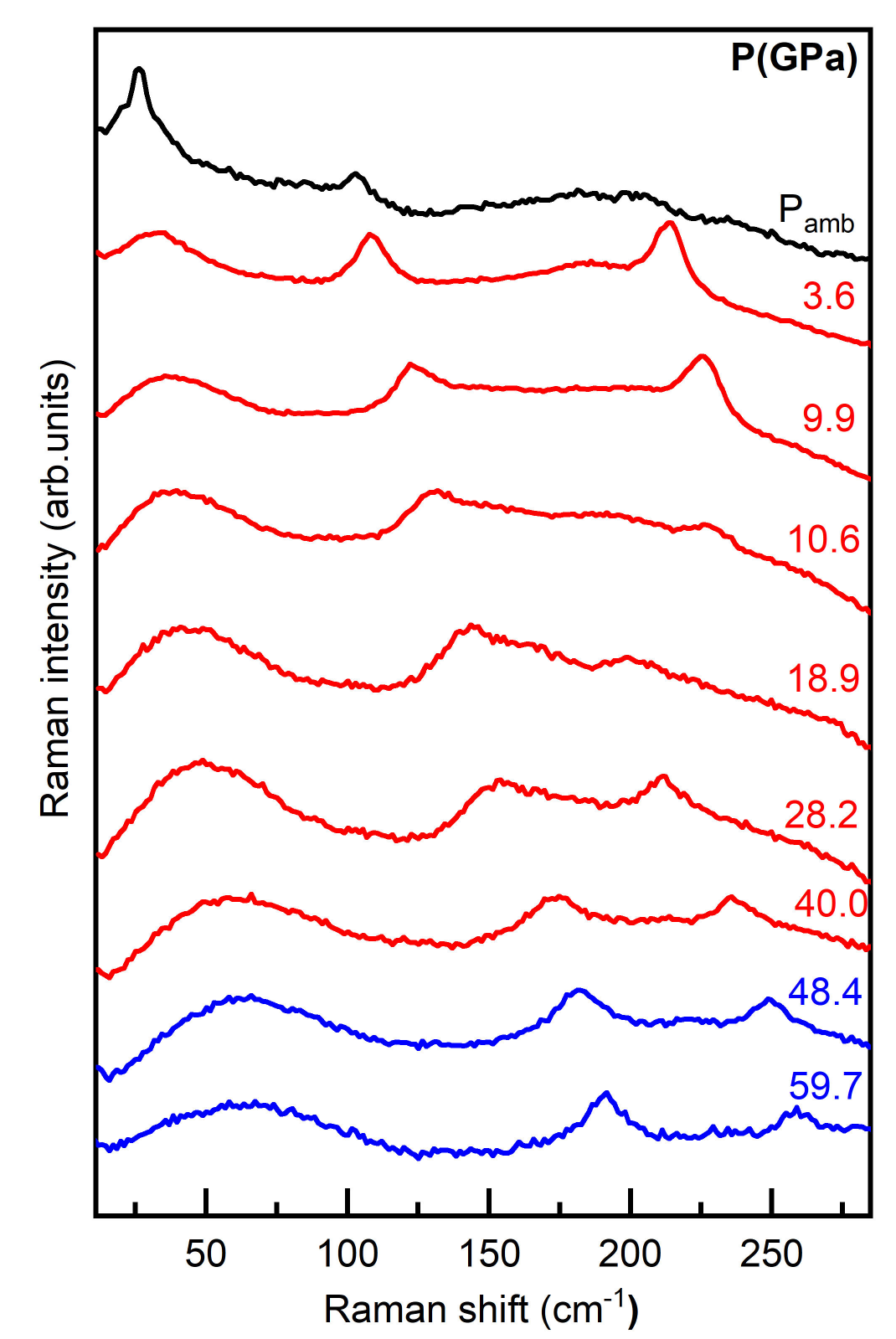}
    \caption{Selected high pressure Raman  spectra upon pressure release from 59.7GPa.}
\end{figure}

\begin{figure}
    \centering
    \includegraphics[width=0.5\linewidth]{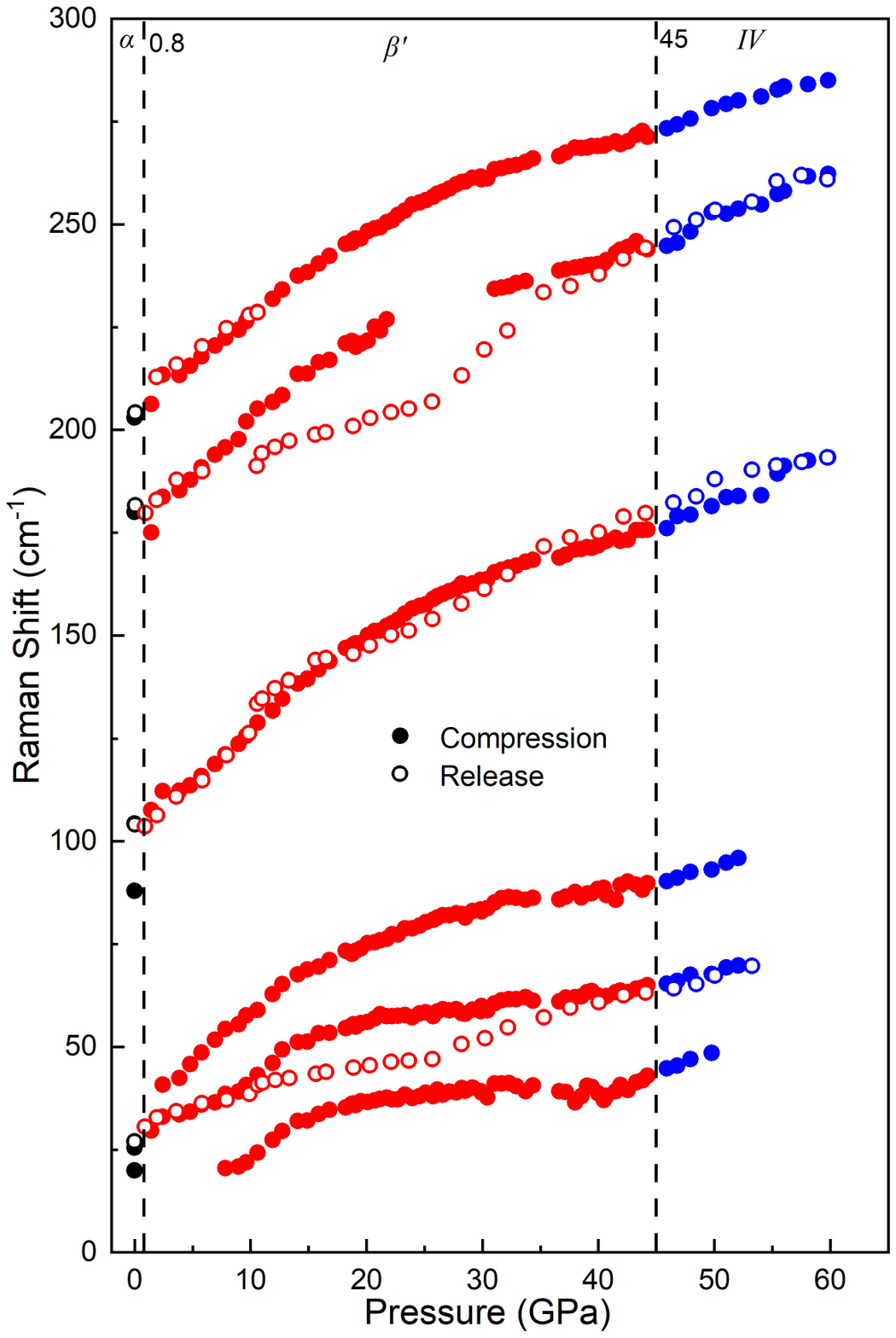}
    \caption{Pressure dependence of the  Raman-active mode frequencies upon pressure increase (solid symbols) and release (open circles).  The $\alpha$, $\beta^{'}$, and IV phases are indicated  with black, red and  blue symbols, respectively. The vertical dashed lines indicate the critical pressures associated with the phase transitions.}
\end{figure}

\begin{figure}
    \centering
    \includegraphics[width=0.5\linewidth]{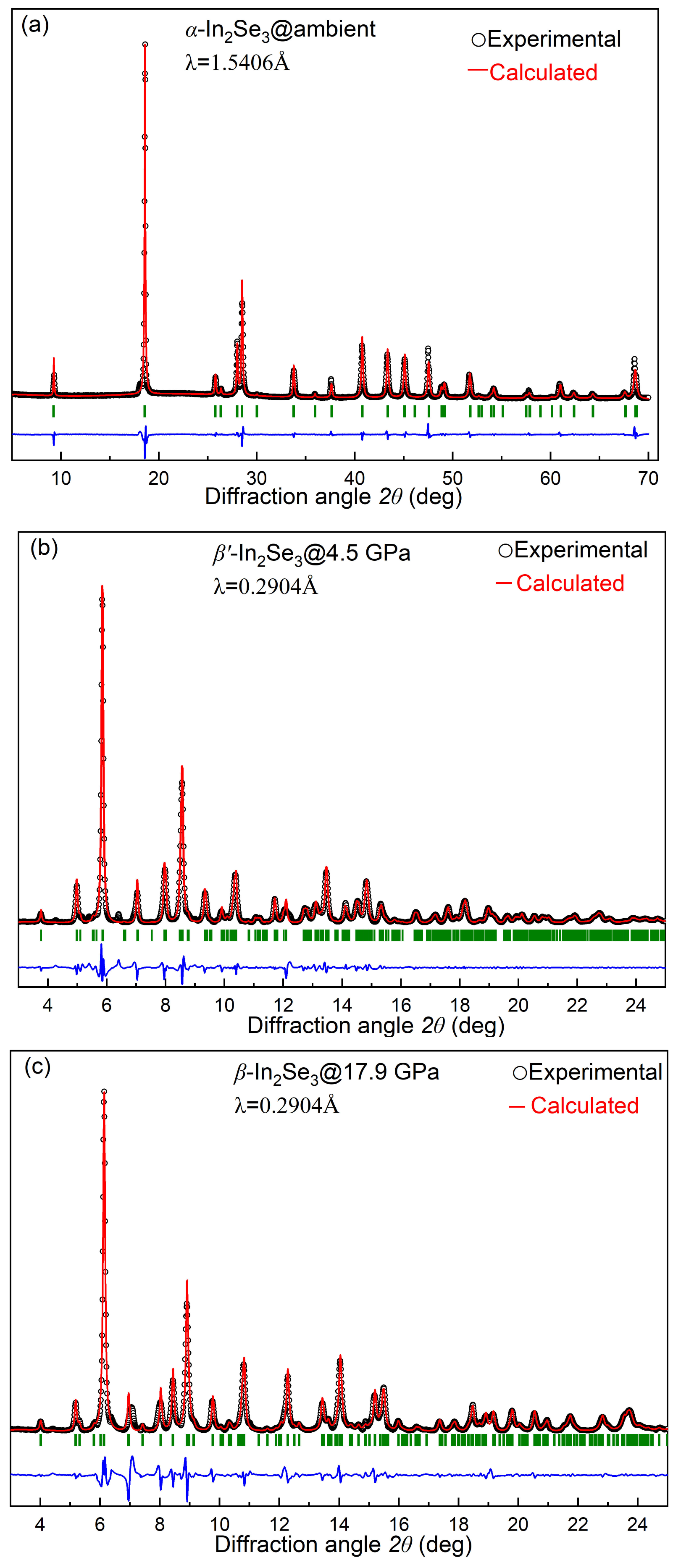}
    \caption{Representative Le Bail refinements for the  (a) $\alpha$-In$_2$Se$_3$,(b)  $\beta$'-In$_2$Se$_3$ and (c) $\beta$-In$_2$Se$_3$  phases.Black open symbols correspond to the measured profile and the red solid lines represent the results of the refinement. The difference curves (blue curves) are shown also. Green vertical ticks mark the positions of the Bragg peaks of the corresponding phases. }
\end{figure}

\begin{figure}[h]
    \centering
    \includegraphics[width=0.5\linewidth]{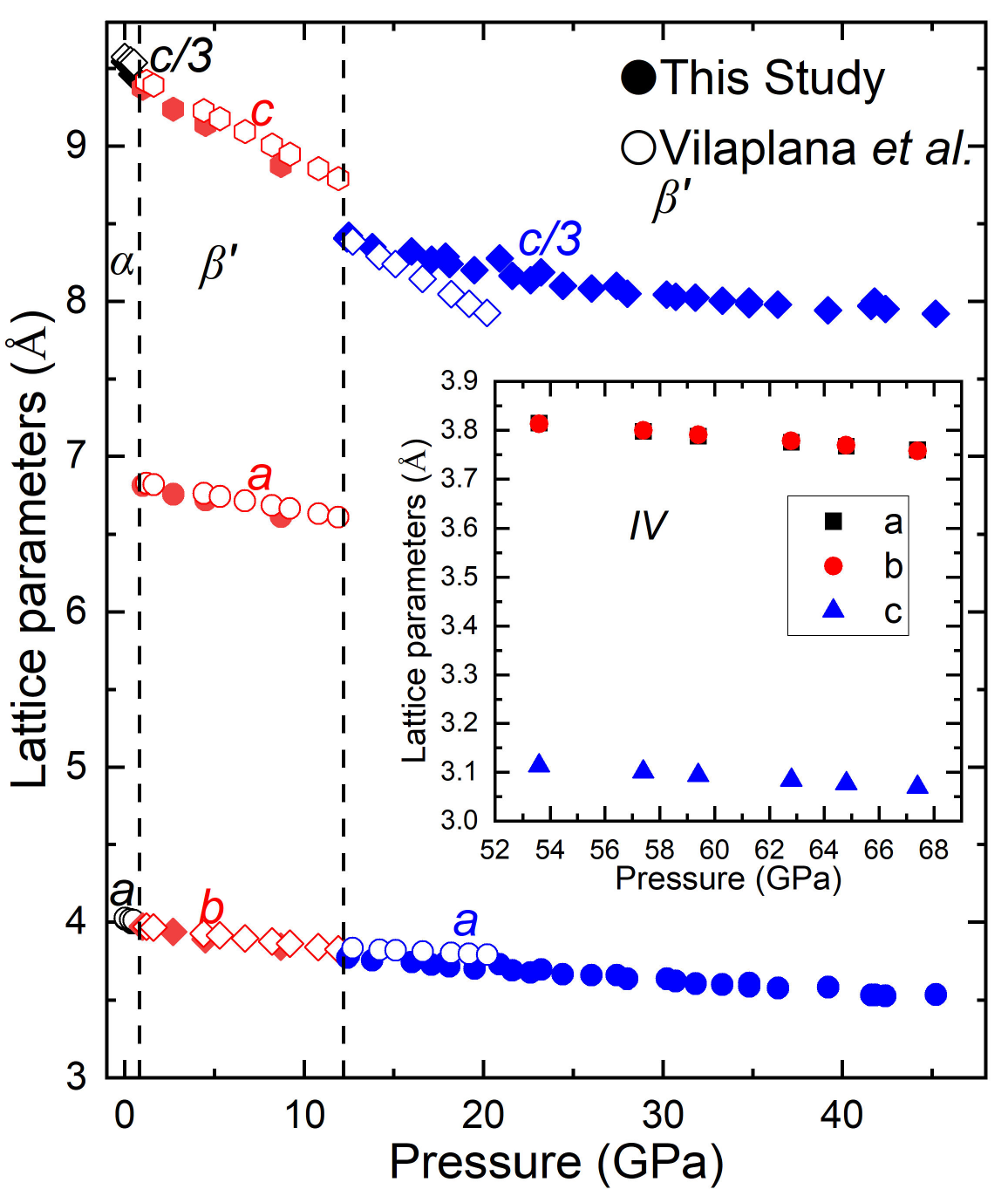}
    \caption{Pressure dependence of the lattice parameters of In$_2$Se$_3$  according to this study  (solid symbols)  and from  Vilaplana  $et$ $al.$ \cite{Vilaplana2018} (open symbols). The inset shows the lattice parameters for the  IV phase. The vertical dashed lines indicate the critical pressures of the  phase transitions.}
\end{figure}

\begin{figure}[h]
    \centering
    \includegraphics[width=0.5\linewidth]{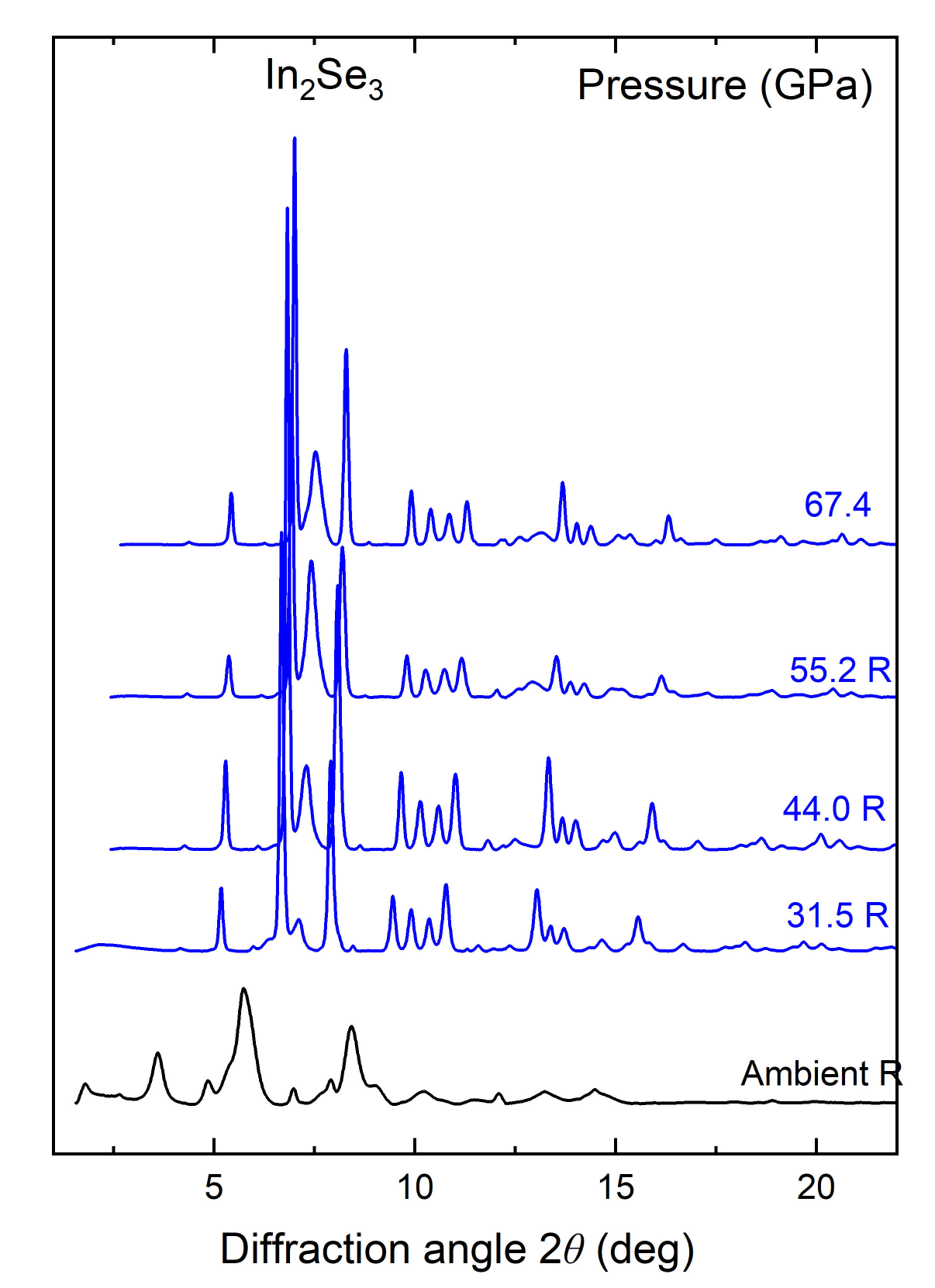}
    \caption{Selected XRD patterns of In$_2$Se$_3$ with decreasing pressure. The $\alpha$-In$_2$Se$_3$,  and IV-In$_2$Se$_3$ phases are denoted with black and  blue colors, respectively. The X-ray wavelength is $\lambda$=0.2904\r{A}}
\end{figure}

\end{document}